\documentclass[11pt,a4paper]{article}
\pdfoutput=1
\usepackage{jheppub}

\usepackage{color}
\usepackage{amsmath, amssymb}
\usepackage{pifont}

\usepackage[utf8]{inputenc}
\usepackage{verbatim}   
\usepackage{amsfonts}
\usepackage{amssymb}
\usepackage{graphicx}
\usepackage{slashed}
\usepackage[normalem]{ulem}

\definecolor{myorange}{RGB}{199,146,32}

\newcommand{\eV}{{\, {\rm eV}}}

\newcommand{\MeV}{{\, {\rm MeV}}}
\newcommand{\GeV}{{\, {\rm GeV}}}
\newcommand{\TeV}{{\, {\rm TeV}}}

\def\beq{\begin{equation}}
\def\eeq{\end{equation}}
\def\bea{\begin{eqnarray}}
\def\eea{\end{eqnarray}}
\def\bitem{\begin{itemize}}
\def\eitem{\end{itemize}}
\newcommand{\bec}{\begin{center}}
\newcommand{\eec}{\end{center}}
\newcommand{\ba}{\begin{array}}
\newcommand{\ea}{\end{array}}

\def\inv{^{\raise.15ex\hbox{${\scriptscriptstyle -}$}\kern-.05em 1}}
\def\lbar{{\lower.35ex\hbox{$\mathchar'26$}\mkern-10mu\lambda}} 

\let\<=\langle
\let\>=\rangle

\let\+=\uparrow

\begin{document}

\begin{flushright}
KCL-PH-TH/2019-12
\end{flushright}
\title{Hearing without seeing: gravitational waves from hot and cold hidden sectors}
\author[a]{Malcolm Fairbairn,}
\author[b]{Edward Hardy,}
\author[a]{Alastair Wickens}
\emailAdd{malcolm.fairbairn@kcl.ac.uk}
\emailAdd{ehardy@liverpool.ac.uk}
\emailAdd{alastair.wickens@kcl.ac.uk}
\affiliation[a]{Department of Physics, King’s College London, Strand, London WC2R 2LS, United Kingdom}
\affiliation[b]{Department of Mathematical Sciences, University of Liverpool, Liverpool L69 7ZL, United Kingdom}

\abstract{
We study the spectrum of gravitational waves produced by a first order phase transition in a hidden sector that is colder than the visible sector. In this scenario, bubbles of the hidden sector vacuum can be nucleated through either thermal fluctuations or quantum tunnelling. If a cold hidden sector undergoes a thermally induced transition, the amplitude of the gravitational wave signal produced will be suppressed and its peak frequency shifted compared to if the hidden and visible sector temperatures were equal. This could lead to signals in a frequency range that would otherwise be ruled out by constraints from big bang nucleosynthesis. Alternatively, a sufficiently cold hidden sector could fail to undergo a thermal transition and subsequently transition through the nucleation of bubbles by quantum tunnelling. In this case the bubble walls might accelerate with completely negligible friction. The resulting gravitational wave spectrum has a characteristic frequency dependence, which may allow such cold hidden sectors to be distinguished from models in which the hidden and visible sector temperatures are similar. We compare our results to the sensitivity of the future gravitational wave experimental programme. 
}

\maketitle


\section{Introduction}

A new hidden sector is a well motivated extension of the Standard Model (SM), both from top down considerations of string theory \cite{Acharya:2017kfi} and from a phenomenological perspective e.g. because it could contain the dark matter (DM) \cite{Schwaller:2015tja}. However, any hidden sector present might be extremely weakly coupled to the visible sector, in which case it would be challenging or even impossible to directly probe \cite{Fairbairn:2018bsw}.\footnote{For small, but not too small, couplings a hidden sector might be detected at beam dump or precision laboratory experiments. Existing bounds from these sources already exclude hidden sectors at low mass scales unless they are fairly weakly coupled to the visible sector.} One signal that could be observed even in the limit of a vanishing coupling to the SM is a background of gravitational waves left over from a phase transition in the hidden sector which occurred early in the universe's cosmological history. 
This is a worthwhile possibility to explore, since the sensitivity and frequency coverage of experimental searches for gravitational waves will improve dramatically in the near future as new instruments are developed and deployed, even though not all hidden sector models have a phase transition that leads to such a signal.

If a hidden sector is extremely weakly coupled to the visible sector there is no reason to expect that the two should be at the same temperature in the early universe. Indeed, as we will discuss, if a hidden sector contains relatively light degrees of freedom it might need to be cold to be compatible with constraints on the effective number of additional relativistic degrees of freedom at the time of big bang nucleosynthesis (BBN) and at the formation of the cosmic microwave background (CMB). A hidden sector that contains stable heavy states without efficient annihilation channels may also need to be cold to avoid these over-closing the universe (e.g. this is the case for pure gauge hidden sectors that contain stable glueballs).

If a phase transition in the early universe is first order, it can release an observationally significant quantity of energy into gravitational waves. 
In this case the transition proceeds by the nucleation and expansion of bubbles of the low temperature phase.  In contrast, a second order phase transition happens smoothly and does not lead to significant gravitational wave emission.

In this paper we study the nature of phase transitions in hidden sectors and the resulting gravitational wave signals, focusing on the impact of a hidden sector being cold relative to the visible sector. Our aim is to make progress towards determining the phenomenological possibilities in generic hidden sectors and understanding what could be inferred about the source of a gravitational wave signal were one to be detected. However, given the challenges involved in analysing phase transitions, we focus on a simple class of hidden sectors that can possess many of the features of interest.

There are multiple dynamical processes that can produce gravitational waves during first order phase transitions. We will argue that the dominant production mechanisms can differ in hot and cold hidden sectors and that this may lead to observable differences in the spectrum emitted. We will also show that a cold hidden sector could lead to gravitational wave signals in a frequency range that would otherwise not be possible owing to BBN constraints (this has recently also been studied in \cite{Breitbach:2018ddu}).

Phase transitions in the early universe and their resulting gravitational wave signals have been a subject of extensive previous study, and we will draw on many results from the literature. Previous work includes hydrodynamical analysis of the motion of bubble walls \cite{Enqvist:1991xw,Espinosa:2010hh,Leitao:2015ola}, studies of the possible extensions of the SM that might lead to gravitational wave signals \cite{Nardini:2007me,Espinosa:2007qk,Espinosa:2008kw,Schwaller:2015tja,Addazi:2016fbj,Jaeckel:2016jlh,Dev:2016feu,Addazi:2017gpt,Aoki:2017aws,Baldes:2017rcu,Ellis:2018mja,Baldes:2018emh,Brdar:2018num} and the possibilities for model discrimination \cite{Jinno:2017ixd,Croon:2018erz}. There have also been extensive theoretical studies of the friction experienced by bubble walls due to a thermal bath \cite{Bodeker:2009qy,Bodeker:2017cim,Moore:1995si,Moore:1995ua,Moore:2000jw,John:2000zq,Kozaczuk:2015owa}, numerical \cite{Huber:2008hg,Hindmarsh:2013xza,Hindmarsh:2015qta,Weir:2016tov,Konstandin:2017sat,Hindmarsh:2017gnf,Cutting:2018tjt} and analytical \cite{Kosowsky:1992rz,Kamionkowski:1993fg,Ignatius:1993qn,Kosowsky:2001xp,Hindmarsh:2016lnk,Jinno:2016vai,Jinno:2017fby} studies of the dynamics of the expanding bubbles and the resulting gravitational signals, as well as analysis of the sensitivity of upcoming experiments. A recent review, containing many further important references, can be found in \cite{Binetruy:2012ze}.

Turning to the structure of this paper: in Section~\ref{sec:model} we describe our example hidden sector, and the mechanisms and conditions for it to be cold relative to the visible sector. In Section~\ref{sec:cosmo} we study the constraints on such a hidden sector from cosmology. In Section~\ref{sec:phaset} we analyse the nature of the different phase transitions possible in such hidden sectors. In Section~\ref{sec:wallv} we study the friction experienced by bubble walls, and their resulting velocities, in different types of transitions.  
In Section~\ref{sec:gvsignals} we discuss the resulting gravitational wave signals in detail and analyse the possibilities for detection and model discrimination in future experiments. Finally, we highlight the remaining uncertainties and areas for future work and conclude in Section~\ref{sec:conc}.

\section{An example hidden sector}\label{sec:model}

Predicting the gravitational wave signal from a particular hidden sector is challenging. It can be hard to reliably determine if a theory has a first order phase transition at all, and it is harder still to obtain enough information about the finite temperature effective potential to calculate the rate at which bubbles of the true vacuum are nucleated as a function of temperature. The friction on the bubble walls from the hidden sector thermal bath must also be calculated and the dynamics of the bubble walls and plasma analysed, which remains a major source of theoretical uncertainty even in the most heavily studied models. We therefore focus on a simple class of hidden sectors in which perturbative calculations are possible, at least to the degree of accuracy required for our present work.

Our example hidden sector consists of an SU(2) gauge group with coupling constant $g$ and a dark scalar field  $\Phi$, the ``hidden sector Higgs'', which is in the fundamental of the gauge group and has a tree level potential
\beq \label{eq:phipot}
V = -m^2 \left|\Phi \right|^2 + \lambda \left|\Phi \right|^4 ~.
\eeq
A similar hidden sector, albeit with large couplings to the visible sector, has been studied in the context of baryogenesis \cite{Katz:2016adq}. As we will discuss in Section~\ref{sec:cosmo} such a sector also leads to a viable DM candidate in parts of parameter space, and related models have been considered in \cite{Hambye:2008bq,Hambye:2013sna,Khoze:2014xha,Karam:2015jta,Baldes:2018emh, Hambye:2018qjv}. The gravitational wave signals from classically scale invariant hidden sectors (at the same temperature as the visible sector) have also been studied in \cite{Jaeckel:2016jlh}. We consider models that are close to the scale invariant limit and share some features with those in \cite{Jaeckel:2016jlh}.

In some parts of parameter space $\Phi$ gets a vacuum expectation value (VEV). We parameterise $\Phi = \left( 0, \frac{1}{\sqrt{2}} \phi \right) $, where $\phi$ is the hidden sector field that gets a VEV $\left<\phi\right>$, so the resulting hidden sector gauge boson masses are $m_A = \frac{1}{2} g \left<\phi\right>$. 

We consider theories with $g^2 \gg \lambda$ and further assume that the tree level mass squared in Eq.~\eqref{eq:phipot} satisfies $\left| m^2 \right| \ll \left<\phi\right>^2$ (this will be seen to be consistent). In this part of parameter space the 1-loop Coleman-Weinberg potential is comparable to the tree level potential \cite{Coleman:1973jx}. We make this choice because the combination of the tree and 1-loop potentials can lead to a first order phase transition but also to induce a barrier in the potential at zero temperature between a meta-stable vacuum and the true vacuum. Neither of these features are possible if the tree level potential Eq.~\eqref{eq:phipot} dominates. In the regime we consider it is convenient to write the mass squared parameter in Eq.~\eqref{eq:phipot} in terms of a dimensionless parameter $\tilde{m}^2$ and a renormalisation group (RG) scale $w$
\beq
m^2 \equiv \tilde{m}^2 \frac{9 g^4}{1024 \pi^2} w^2 ~.
\eeq
We choose the RG scale to coincide with the VEV $\left<\phi\right> = w$ (in parts of parameter space in which a symmetry breaking vacuum exists).

It is straightforward to evaluate the 1-loop correction to $\phi$'s potential. Given the assumption of a small quartic coupling we can neglect loops of $\phi$ itself, and the result comes only from the hidden sector gauge bosons. After adding appropriate counterterms, the total zero temperature potential is 
\beq \label{eq:v0}
V_0\left(\phi\right) = \frac{9 g^4}{1024 \pi^2} \left[ \frac{1}{2} \tilde{m}^2 w^2 \phi^2 + \phi^4 \left( \log\left(\frac{\phi^2}{w^2}\right)- \frac{(\left(2+\tilde{m}^2 \right)}{4} \right) \right]~.
\eeq
As usual the quartic coupling has been replaced with the renormalisation scale $w$ and renormalisation conditions by dimensional transmutation.\footnote{The contribution to the zero temperature potential from $\phi$ itself is of the form $V = \frac{1}{64 \pi^2} \left( V''(\phi) \right)^2 \log(\frac{V''\left(\phi\right)}{w^2})$. Since $V'' \sim 9 g^4 w^2/(64 \pi^2)$ our analysis is consistent for $g\sim 1$.}

If $\tilde{m}^2\leq 0$ the point $\left<\phi\right> =0$ is unstable at zero temperature, if $0<\tilde{m}^2<2$ it is a metastable minimum, and if $\tilde{m}^2>2$ this is the true vacuum.
If $0<\tilde{m}^2<2$ 
the difference in energy density between the true vacuum at $\left<\phi\right> \neq 0$ and the metastable vacuum at $\left<\phi\right> =0$ at zero temperature is
\beq \label{eq:deltaV}
\rho_{\rm vac} = \frac{9 g^4}{1024 \pi^2}  \frac{\left(2- \tilde{m}^2 \right)}{4} w^4 ~,
\eeq
and the mass of $\phi$ in the symmetry breaking vacuum is
\beq \label{eq:mphi2}
m_{\phi}^2=\frac{9}{512\pi^2} \left( 4-\tilde{m}^2\right) g^4 w^2 ~.
\eeq

Phase transitions in the early universe depend on $\phi$'s potential at finite temperature. The simplest estimate of this is obtained by combining the zero temperature potential, Eq.~\eqref{eq:v0}, with the naive one loop finite temperature correction $V_T$ \cite{Kapusta:2006pm}
\beq
V\left(\phi\right)= V_0\left(\phi\right) + V_T\left(\phi\right)~.
\eeq
The contribution to $V_T\left(\phi\right)$ from the hidden sector gauge bosons is
\beq \label{eq:finiteTpot}
V_T\left(\phi\right) \supset \frac{n_i T^4}{2\pi^2}  \int_0^{\infty} q^2 \log \left(1- \exp\left(-\sqrt{q^2+m_A^2\left(\phi\right)/T^2} \right) \right)~dq ~,
\eeq
where $n_i=9$ and $m_A\left(\phi\right) = \frac{1}{2} g \phi$.  The one loop correction from $\phi$ loops has a similar form and at temperatures around the time of the phase transition is subleading to Eq.~\eqref{eq:finiteTpot}, as is the case with the zero temperature potential.\footnote{This can be seen directly by expanding the integral analogous to that in Eq.~\eqref{eq:finiteTpot}.} 

Although it demonstrates the existence of a phase transition, the simple one-loop thermal potential is known to lead to significant inaccuracies in many models and can even lead to incorrect predictions of the order of a transition. Different approaches have been proposed to capture the effects missed by Eq.~\eqref{eq:finiteTpot} (a recent discussion can be found in \cite{Curtin:2016urg}).  In our present work we are interested in phenomenological possibilities rather than precise demarcation of the parameter space. It is therefore sufficient to only improve Eq.~\eqref{eq:finiteTpot} by resumming an infinite set of daisy diagrams. This fixes the most severe shortcoming of Eq.~\eqref{eq:finiteTpot} by removing IR divergences that would otherwise spoil the perturbative loop expansion  \cite{Carrington:1991hz,Fendley:1987ef}. In practice the daisy resummation can be performed by simply replacing the masses in Eq.~\eqref{eq:finiteTpot} 
\beq \label{eq:daisy}
m_i^2\left(\phi\right) \rightarrow m_i^2\left(\phi\right) + \Pi_i~,
\eeq
where $\Pi_i$ is the finite temperature self energy of the species $i$. At leading order in $T^2$ the longitudinal components of the hidden sector gauge bosons have $\Pi_{{\rm long}} = \frac{11}{6} g^2 T^2$ and the transverse gauge bosons have no dependence at this order, and $\Pi_{\phi} = \frac{1}{2} \left(g^2 + \lambda \right)T^2$ \cite{Espinosa:1995se}.

\begin{figure}[t]\begin{center}
\includegraphics[height=4.8cm]{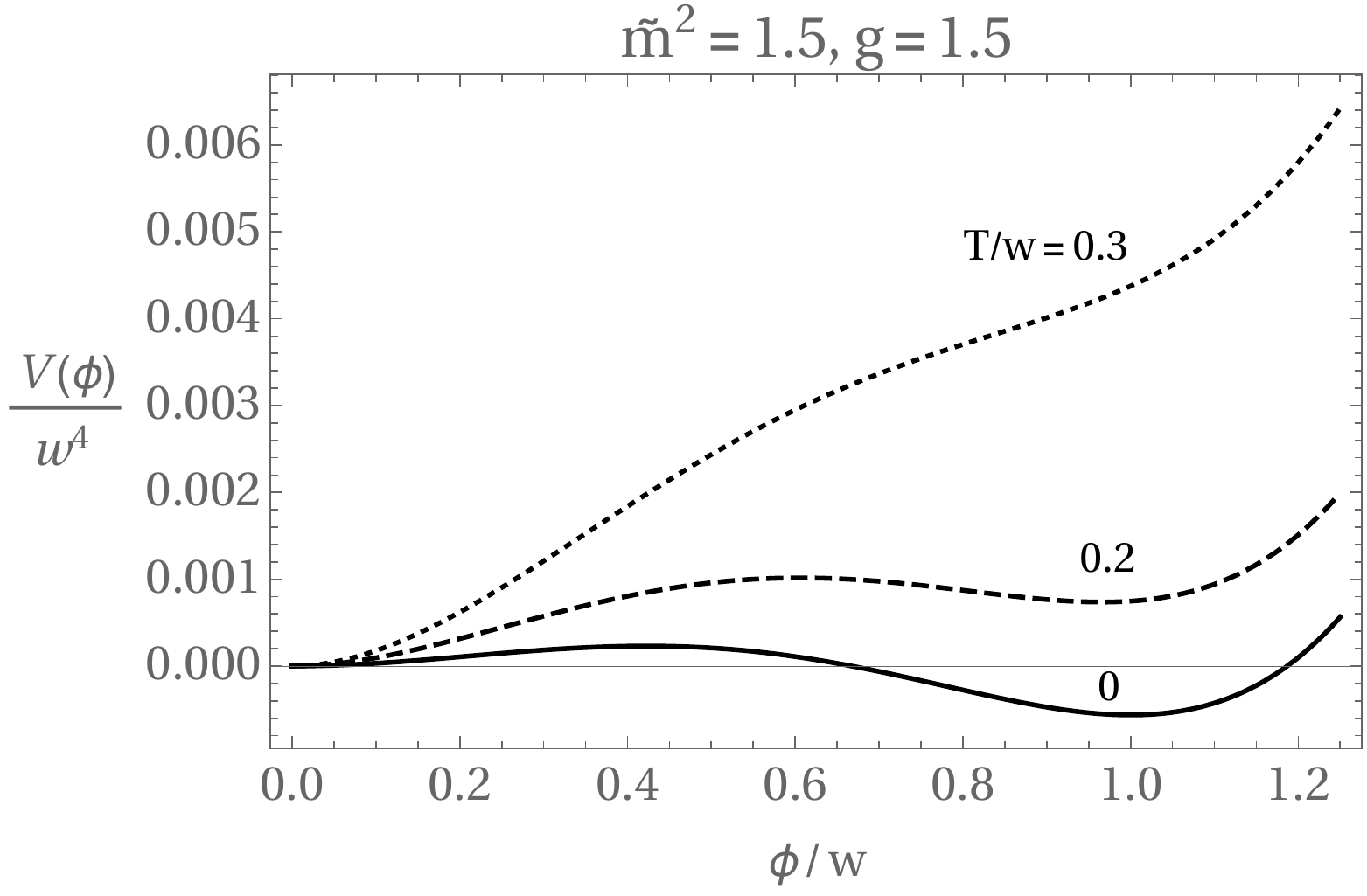}\hspace{0.1cm}
\includegraphics[height=4.8cm]{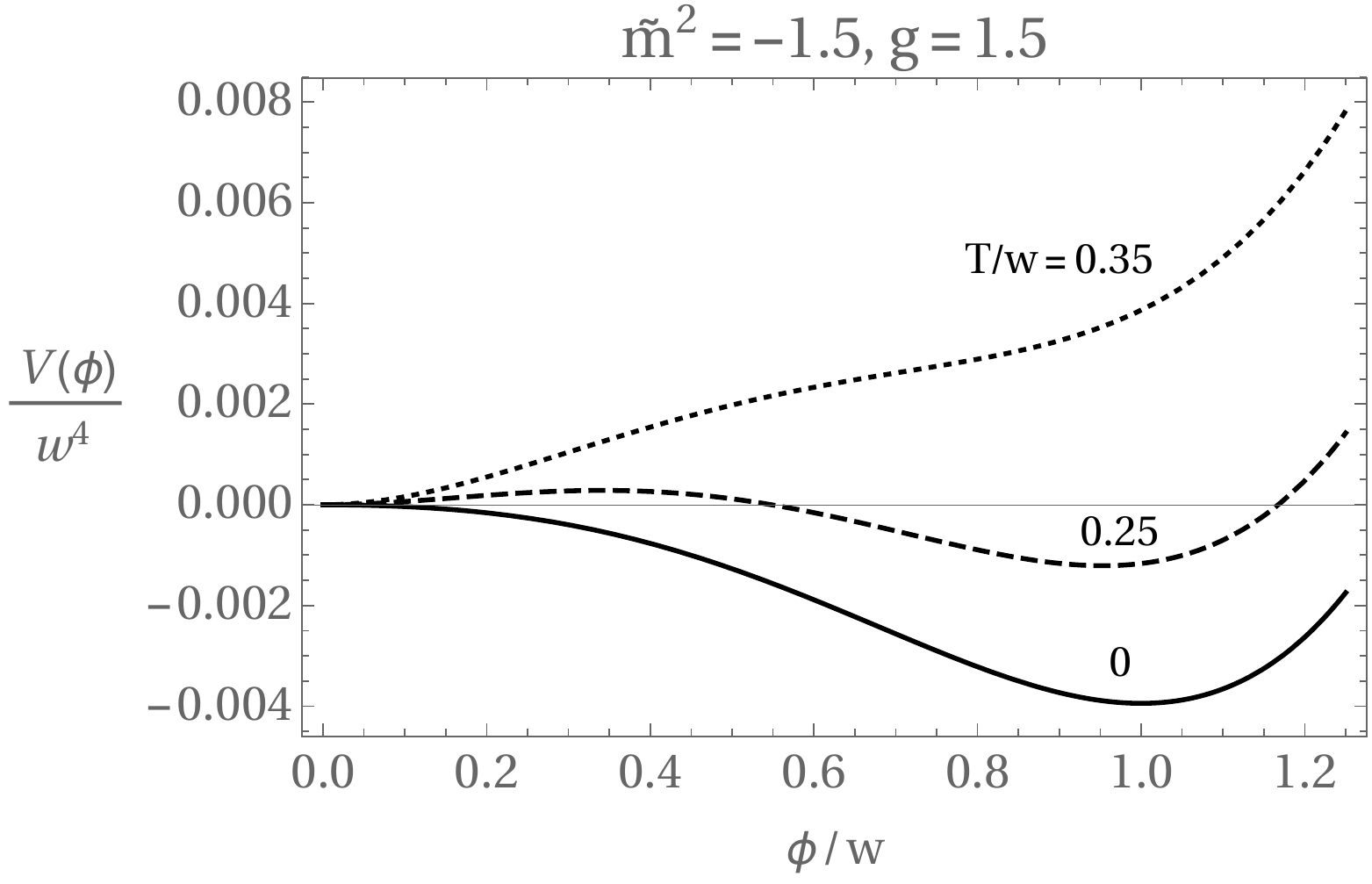} \end{center}
\caption{The scalar potential $V\left(\phi\right)$ of the hidden sector that we consider, at different points in parameter space (left and right), at zero temperature (solid) and increasing temperature (dashed, dotted). The model plotted in the left panel has a barrier between the two vacua that remains at zero temperature, and the barrier disappears at zero temperature in the model in the right panel.}
\label{fig:pot}
\end{figure}

Having made the modification in Eq.~\eqref{eq:daisy} it is straightforward to calculate $\phi$'s finite temperature potential, and we show examples of the results in Figure~\ref{fig:pot}. The RG scale $w$ is the only dimensionful parameter in the hidden sector and fixes the overall scale. At high temperatures $\phi=0$ is always favoured and if this field value remains as a stable minimum there will be no phase transition, but if it persists as a metastable minimum there could be a first order phase transition. Further, if $\phi=0$ becomes a local maximum by the time the temperature reaches zero, the phase transition could be second order if the barrier disappears before the $\phi \neq 0$ minimum has lower energy than $\phi=0$. Conversely the transition could be first order if this happens while a barrier remains.\footnote{In contrast if $\lambda \gtrsim g$ so that the tree-level potential dominates over the 1-loop potential any phase transition will be second order, unless there are additional hidden sector degrees of freedom.}

\subsection{Generating and maintaining a cold hidden sector}

We define the temperature ratio between the hidden and visible sectors to be
\beq \label{eq:epdef}
\epsilon \equiv \frac{T_{\rm h}}{T_{\rm v}} ~,
\eeq
where $T_{\rm h}$ and $T_{\rm v}$ are the hidden and visible sector temperatures at some time prior to the phase transition.

If there is no energy exchange between the visible and hidden sectors and no entropy injection into either sector, $\epsilon$ is approximately constant during the evolution of the universe. In this case it only evolves due to changes in the number of relativistic degrees of freedom in the two sectors and 
\beq \label{eq:changeTg}
\frac{g_{\rm h}  T_{\rm h}^3}{g_{\rm v}  T_{\rm v}^3} = \frac{g_{\rm h, RH} T_{\rm h,RH}^3}{g_{\rm v, RH} T_{\rm v,RH}^3} ~,
\eeq
where $g_{\rm h}$ and $g_{\rm v}$ are the number of relativistic degrees of freedom in the hidden and visible sectors respectively, and ${\rm RH}$ indicates that a quantity is defined immediately after reheating completes. The resulting changes in $\epsilon$ are relatively small and do not matter when considering extreme temperature hierarchies. However, only a mild temperature difference is needed to satisfy observational constraints from BBN and the CMB so the effects of Eq.~\eqref{eq:changeTg} are relevant when considering these bounds.

The universe might enter its final period of radiation domination when the inflaton decays at the end of inflation. Alternatively, in many string theory models the universe goes through a period of matter domination after inflation. This is due to the presence of relatively light and long lived moduli that are initially displaced from the minimum of their potentials, before being reheated for a final time when the longest lived of these decays. In both of the above cases the initial value of $\epsilon$ is set by the partial decay rates to the visible and hidden sectors ($\Gamma_{\rm v}$ and $\Gamma_{\rm h}$, respectively) of the state responsible for reheating the universe for the final time. Then assuming the energy density in the two sectors is negligible prior to reheating, the temperature ratio just after this is
\beq
\epsilon_{\rm RH} = \left(\frac{g_{\rm v} \Gamma_{\rm h}}{g_{\rm h} \Gamma_{\rm v}}  \right)^{1/4} ~,
\eeq
where $g_{\rm v}$ and $g_{\rm h}$ are the effective number of relativistic degrees of freedom in the visible and hidden sectors immediately after reheating. It is plausible that $\Gamma_{\rm h}$ and $\Gamma_{\rm v}$ could differ dramatically leading to a significant temperature ratio. For example, this could occur if the longest lived modulus in a string compactification comes from a cycle associated with the visible sector, and the hidden sector is localised elsewhere.

Having obtained a hierarchy in their initial values, we also require that the temperature ratio between the hidden and visible sectors persists until the hidden sector phase transition occurs. This can be achieved simply by assuming that the two sectors are completely decoupled.\footnote{There is still the possibility of thermalisation via the inflaton if this has relatively large couplings, however if it decays via non-renormalisable operators this effect is negligible \cite{Adshead:2016xxj,Hardy:2017wkr}.} 

On the other hand, it is interesting to consider the size of interactions between the two sectors that are allowed without the temperature hierarchy being destroyed. The presence of such a coupling would affect the cosmological history of a hidden sector, for example by allowing otherwise stable hidden sector states to decay, and could potentially also lead to observable signals of a hidden sector.

In Appendix \ref{app:maintaing-temp} we show that maintaining a large temperature hierarchy requires parameterically smaller portal couplings than the well known conditions for a hidden sector to remain out of thermal equilibrium with the visible sector. For example, the constraint on the Higgs portal coupling,
\beq \label{eq:hpop}
\mathcal{L} \supset -\frac{1}{2}\lambda_p \left|\Phi\right|^2 \left|H \right|^2 ~,
\eeq
where $H$ is the SM Higgs doublet, to maintain a temperature hierarchy is found to be
\beq \label{eq:cond1}
\lambda_p \lesssim 10^{-10} \epsilon^2 ~.
\eeq

\section{Cosmological constraints on non-thermalised sectors}\label{sec:cosmo}

A hidden sector can affect the cosmological history of the universe, and the requirement that its effects do not lead to contradictions with observations can exclude large regions of parameter space. In particular, the energy density in a hidden sector should not destroy the successful predictions of BBN or leave an imprint in the CMB, while any stable relics that it contains must not overclose the universe. The resulting constraints depend on when the hidden sector phase transition happens relative to events in the visible sector. In Section~\ref{sec:phaset} we will see that hidden sector phase transitions can take place at hidden sector temperatures $T_{\rm h} \sim w$ or when the hidden sector is much colder $T_{\rm h} \ll w$. Cosmological constraints can arise due to energy in the hidden sector prior to the phase transition, or due to the energy released by the phase transition into the hidden sector.

The class of hidden sectors that we consider has many potentially viable parts of parameter space, and rather than fully explore all of these we simply argue that cosmologically acceptable models can easily be found. For simplicity, we consider theories in which the hidden sector Higgs is lighter than the gauge bosons $m_{\phi} < m_{A}$ (c.f. Eq.~\eqref{eq:mphi2}). In this case there are no decay or annihilation channels for $\phi$ unless it has a coupling to the visible sector. Further, the hidden sector gauge bosons are stable through the analogue of the SM's custodial symmetry. Both $\phi$ and the gauge bosons can be made unstable by introducing new light hidden sector fermions, and we will see that this is necessary to evade cosmological bounds in some parts of parameter space.

We define a parameter characterising the amount of energy released by the hidden sector phase transition 
\beq \label{eq:alpha}
\alpha = \frac{\rho_{\rm vac}}{\rho_{\rm v}}~,
\eeq
where $\rho_{\rm vac}$ is the energy density released by the phase transition, and $\rho_{\rm v}$ is the energy density in the visible sector thermal bath when it occurs. 
We also introduce a similar parameter measuring the energy released relative to that in the hidden sector thermal bath, $\rho_{\rm h}$, immediately prior to the transition 
\beq
\alpha_{\rm h} = \frac{\rho_{\rm vac}}{\rho_{\rm h}}~.
\eeq
We will see in Section~\ref{sec:phaset} that $\alpha \lesssim 10$ and both $\alpha_{\rm h} \sim 1$ and $\alpha_{\rm h} \gg 1$ are possible, depending on the type of transition.

\subsection{Constraints from stable relics}

First we estimate the relic abundance of $\phi$ if it is stable. Assuming that its number changing interactions become inefficient immediately after the hidden sector phase transition, the relic yield of $\phi$  is
\beq
\begin{split}
Y_{\phi} &= \frac{n_{\phi}}{s_{\rm v}} ~,
\end{split}
\eeq
where $s_{\rm v}$ is the visible sector entropy density at this time (which is assumed to dominate the universe). 
If the energy released by the phase transition is small compared to the energy in the hidden sector at this time, i.e. $\alpha_{\rm h}\lesssim 1$, the relic yield is
\beq \label{eq:phirel}
\begin{split}
Y_{\phi}&\sim \frac{80}{g_{\rm v} g^2} \epsilon^3 \\
\end{split}
\eeq
where $g_{\rm v}$ and $\epsilon$ are evaluated at the time of the phase transition. For the relic population of $\phi$ not to exceed the observed DM abundance 
\beq
Y_{\phi} < \frac{4.4 \times 10^{-10} ~ \GeV}{m_{\phi}}~,
\eeq
which constrains
\beq \label{eq:epconstPHI}
\epsilon \lesssim 2 \times 10^{-3} \left(\frac{\GeV}{w}\right)^{1/3} \left(\frac{g_{\rm v}}{106.75} \right)^{1/3} ~,
\eeq
at the time of the phase transition.\footnote{We assume that the hidden sector is in internal thermal equilibrium immediately before it undergoes its phase transition. This will be the case unless there is an extreme hierarchy between the hidden and visible sectors' temperatures, which would result in a negligibly small relic abundance from energy in the hidden sector prior to the transition.}

If the energy released by the phase transition is large compared to that previously in the hidden sector thermal bath, the relic abundance of $\phi$ constrains $\alpha$ rather than $\epsilon$. Since the hidden sector is reheated to a temperature $T_{\rm h}\lesssim w$, the number of $\phi$ states produced is approximately $\rho_{\rm vac}/m_{\phi}$ and the yield of $\phi$ can be estimated as
\beq \label{eq:phialpye}
Y_{\phi} \sim \frac{1}{g_{\rm v}^{1/4}} \alpha^{3/4} ~.
\eeq
For this to not over close the universe requires\footnote{If the energy in the hidden sector thermal bath immediately before the transition is similar to that released, so $\alpha_{\rm h}\sim 1$, the constraints Eq.~\eqref{eq:epconstPHI} and Eq.~\eqref{eq:alphaRelicLim} coincide, as expected.}
\beq \label{eq:alphaRelicLim}
\alpha \lesssim 10^{-11} \left(\frac{g_{\rm v}}{106.75}\right)^{1/3} \left(\frac{\GeV}{g^2 w} \right)^{2/3} ~.
\eeq

Although they are only very approximate, Eq.~\eqref{eq:epconstPHI} and \eqref{eq:alphaRelicLim} are sufficient to show that if $\phi$ is stable then values of $\epsilon \sim 1$ are not viable regardless of how little energy is released by the hidden sector phase transition. If $\epsilon \ll 10^{-3}$ and $\alpha \ll 10^{-11}$ there is no danger of a relic population of $\phi$ over closing the universe. However, we will see that the gravitational wave signals from a sector satisfying these conditions are unobservably small, so we now consider models in which $\phi$ is unstable.

If the Higgs portal operator Eq.~\eqref{eq:hpop} is present $\phi$ can decay to the visible sector once it has a VEV. Its decay rate to a pair of visible sector fermions with mass $m_f$ (after the EW phase transition) is
\beq
\Gamma_{\phi} = \frac{\lambda_p^2}{4\pi} \frac{w^2 m_f^2 m_{\phi}}{m_h^4} \left(1-\frac{4 m_f^2}{m_{\phi}^2}\right)^{3/2} ~,  
\eeq
which, considering visible sector fermions with $m_f \ll m_{\phi}$, corresponds to a lifetime
\beq \label{eq:gammaferm}
\Gamma_{\phi}^{-1} \simeq \left(\frac{10^{-7}}{\lambda_p} \right)^{2} \left(\frac{\GeV^2}{m_f w}\right)^2 {\rm s} ~.
\eeq
If $\phi$ is sufficiently heavy it can also decay directly to the SM Higgs. This is possible while the SM is in the unbroken EW phase (due to the Higgs thermal mass this still requires $m_{\phi}> 2m_h$), but for our purposes it is enough to note that its decay rate once EW symmetry is broken, assuming $m_{\phi}\gg m_{h}$, is 
\beq
\Gamma_{\phi} \simeq \frac{\lambda_p^2 w^2 }{32 \pi m_{\phi}} ~,
\eeq
which leads to a lifetime
\beq \label{eq:gammahp}
\Gamma^{-1}_{\phi} \simeq \left(\frac{10^{-12}}{\lambda_p}\right)^2 \left(\frac{100~\GeV}{w} \right) {\rm s}  ~.
\eeq

In some parts of parameter space Eqs.~\eqref{eq:gammaferm} and \eqref{eq:gammahp} allow $\phi$ to decay before BBN for values of $\lambda_p$ that do not destroy the temperature hierarchy between the hidden and visible sectors. However, this is not possible if $\phi$ is relatively light. In such cases, the simplest option to obtain a viable model is to introduce light hidden sector fermions, with mass $m_{\psi}$, that $\phi$ can decay to. We assume that the coupling of $\phi$ to these states is sufficiently large that it decays fairly fast, but that these states only interact with each other weakly and after the phase transition their comoving number density is constant. If $\alpha_{\rm h} \gtrsim 1$ their yield can be estimated similarly to Eq.~\eqref{eq:phialpye} and to avoid overclosure of the universe requires
\beq \label{eq:mpsilim}
m_{\psi} \lesssim \frac{\eV}{\alpha^{3/4}}~,
\eeq
while if $\alpha_{\rm h}\lesssim 1$ we need
\beq
m_{\psi} \lesssim \frac{\eV}{\epsilon^{3}}~,
\eeq
similarly to how Eq.~\eqref{eq:phirel} was derived.
If $\alpha \sim 1$ the relic population of $\psi$ forms a warm dark matter component, which must be subdominant to the main cold dark matter.

There can also be a significant relic abundance of the hidden sector gauge bosons. Given our assumption about the hierarchy of masses in the hidden sector these have an annihilation channel to $\phi$ (with a cross section that is proportional to $g^2/w^2$), so their relic abundance is set by freeze-out. 

If the hidden sector is at approximately the same temperature as the visible sector, the gauge boson relic abundance is the same as has been studied in the literature. In this case there are large regions of parameter space that have either an under-abundance of gauge boson dark matter, or the full required abundance (provided that $w \lesssim 10^5 ~\GeV$ due to the usual unitarity bound) \cite{Hambye:2008bq}. 
If the hidden sector is cold relative to the visible sector, this will affect the gauge boson relic abundance. The Hubble parameter will be larger when the hidden sector temperature drops below $m_A$, owing to the energy density in the visible sector, so freeze-out will happen at a slightly higher hidden sector temperature than would otherwise be the case. On the other hand, the final dark matter yield will be dramatically decreased since the entropy of the universe is higher, which typically more than compensates the previous effect.

An upper bound on the gauge boson yield can be obtained similarly to that of $\phi$ in Eq.~\eqref{eq:phirel} and \eqref{eq:phialpye}. This corresponds to assuming that gauge boson annihilations become inefficiently immediately after the phase transition. Since we consider parts of parameter space in which the gauge bosons masses are similar to the mass of $\phi$, the resulting bounds on $\alpha$ and $\epsilon$ are similar to Eqs.~\eqref{eq:epconstPHI} and \eqref{eq:alphaRelicLim}. 

We therefore conclude that if $\alpha$ and $\epsilon$ are such that a population of stable $\phi$ particles is cosmologically safe, the hidden sector gauge bosons will also not overclose the universe. Models in which $\phi$ can decay to the visible sector via a Higgs portal operator are only possible with not too different hidden and visible sector temperatures (otherwise the temperature hierarchy would be erased), so there are large regions of parameter space such that the gauge boson relic abundance is viable. Finally in models that require light hidden sector fermions for $\phi$ to decay to, these states can break the hidden sector custodial symmetry, and therefore allow the gauge bosons to also decay.

\subsection{Bounds on additional relativistic degrees of freedom}

The current observational constraint on the effective number of additional relativistic degrees of freedom $g_{\rm new}$ at the time of BBN is (at 95\% confidence level) \cite{Pitrou:2018cgg}
\beq \label{bbn-bound-1}
g_{\rm new} < 0.263~,
\eeq
which, following \cite{Feng:2008mu}, can be interpreted as a constraint on cold hidden sectors. A hidden sector that contains $g_{\rm h, BBN}$ relativistic degrees of freedom with a temperature $T_{\rm h, BBN}$ at the time of BBN gives a contribution to $g_{\rm new}$ of
\begin{align} \label{bbn-hidden-dof}
\begin{split}
\Delta g_{\rm new}  &=  g_{\rm h, BBN} \left(\frac{T_{\rm h, BBN}}{T_{\rm v, BBN}}\right)^4 \\
&= g_{\rm h, BBN}~ \epsilon_{\rm BBN}^4
\end{split}
\end{align}
where ${\rm BBN}$ denotes quantities evaluated at the time of BBN.\footnote{It is straightforward to extend these expressions to include hidden sector states that are on the threshold of becoming non-relativistic.}

If the hidden sector phase transition happens after BBN (so the hidden sector Higgs and gauge fields are relativistic at this time), Eq.~\eqref{bbn-hidden-dof} constrains
\beq
\epsilon_{\rm BBN} < 0.40 ~,
\eeq
assuming there are no additional hidden sector degrees of freedom. If an additional hidden sector Dirac fermion is introduced so that $\phi$ can decay the constraint is slightly stronger
\beq
\epsilon_{\rm BBN} < 0.37 ~.
\eeq
Assuming that entropy is conserved between reheating and the hidden sector phase transition, these can be converted into bounds on $\epsilon$ at the time of reheating via Eq.~\eqref{eq:changeTg}
\beq
\begin{aligned}
\epsilon_{\rm BBN} &\simeq 0.47 \epsilon_{\rm RH} \left(\frac{106.75}{g_{\rm v,RH}} \right)^{1/3} ~\\
\implies  \epsilon_{\rm RH} &<  0.76 \left(\frac{106.75}{g_{\rm v,RH}} \right)^{1/3} ~,
\end{aligned}
\eeq
where we assume the presence of light hidden sector fermions in the last line.

If the phase transition happens before BBN and $\phi$ is stable, the relic density constraints, Eqs.~\eqref{eq:epconstPHI} and \eqref{eq:alphaRelicLim}, are much stronger than BBN bounds so the latter are automatically satisfied in viable models. If $\phi$ decays to the visible sector prior to BBN there are also no constraints from this source.\footnote{Models in which there are decays of hidden sector states to the visible sector around the time of BBN, or subsequently, are strongly constrained \cite{Fradette:2017sdd}, however we focus on parts of parameter space safely away from this issue.}

If the phase transition happens prior to BBN and there are light hidden sector fermions, the constraints from BBN can be significant. Shortly after the phase transition all of the hidden sector's energy density is transferred to a population of $\psi$, and the relic density bound requires that the mass of these is such that they remain relativistic until the time of BBN (assuming a relatively large $\alpha$ for an observable signal). As a result we require
\beq
\begin{aligned}
\alpha &< 0.024 ~,
\end{aligned}
\eeq
if $\alpha_{\rm h} \gtrsim 1$, and $\epsilon < 0.23$ at the time of the phase transition otherwise.

CMB bounds on the additional number of relativistic degrees of freedom might also be relevant. Analogously to Eq.~\eqref{bbn-hidden-dof}, these can be converted to a constraint on a cold hidden sector
\beq \label{cmb-bound}
g_{\rm h, CMB}~ \epsilon_{\rm CMB}^4 \leq 0.0518 \enspace (95\% \enspace \rm CL) ~,
\eeq
at the time of photon decoupling \cite{Aghanim:2018eyx}. 

We assume that the hidden sector phase transition happens long before the formation of the CMB. In this case CMB limits are only important if the hidden sector contains light fermions. We assume $\alpha$ is not too small (as is required for observable gravitational wave signals), so the relic $\psi$ states are still relativistic at CMB time from Eq~.\eqref{eq:mpsilim}. Then Eq.~\eqref{cmb-bound} is satisfied if 
\beq
\alpha < 0.015 ~,
\eeq
and 
\beq
\epsilon < 0.2 ~,
\eeq
at the time of the phase transition. These limits are slightly, but not dramatically, stronger than the  constraints from BBN.

The strong dependence of Eq.~\eqref{bbn-hidden-dof} on $\epsilon$ means that only a mild temperature hierarchy between the hidden and visible sectors is needed to accommodate BBN and CMB observations. Despite this, the temperature difference resulting from the large change in the visible sector number of relativistic degrees of freedom at the QCD phase transition and electron/positron annihilation is not quite sufficient for the constraints to be satisfied if the hidden and visible sectors are initially at the same temperature, and an initial temperature difference is required.\footnote{We assume the SM high temperature value of $g_{\rm v}$ at reheating.}

\subsection{The viable parameter space}

We identify two regions of our model's parameter space that are cosmologically safe and in which it is out of thermal equilibrium with the visible sector. These serve as a basis for our subsequent study of hidden sector phase transitions and their gravitational wave signals.

In the first region, the hidden sector is at a relatively high scale such that $m_{\phi} \gg m_h$ and its temperature is not too different to that of the visible sector, so the hidden sector phase transition happens before EW symmetry breaking. The introduction of a small Higgs portal coupling allows $\phi$ to decay to the visible sector before BBN, and this does not destroy the temperature asymmetry provided $\epsilon \gtrsim 10^{-3}$. Meanwhile, since $\epsilon$ is not too small the annihilation of hidden sector gauge bosons is reasonably efficient compared to the Hubble parameter at the time of the phase transition and the relic abundance of these can easily be viable.

A second possibility is that the hidden sector has no portal coupling to the visible sector. In this case, by introducing light hidden sector fermions, hidden sector models at any scale are viable provided the values of $\epsilon$ and $\alpha$ are such that BBN and CMB constraints are satisfied (assuming the hidden sector fermions are sufficiently light that their relic abundance is small).

We compare the phase transitions that happen in cold hidden sectors to those in hidden sectors that are at the same temperature as the visible sector. It is therefore useful to briefly discuss the parameter space in which such sectors are not excluded.

First we note that any hidden sector at the same temperature as the visible sector is excluded by BBN constraints if it goes through a phase transition at a temperature $T \lesssim 10~\MeV$. For the hidden sector that we consider, this is the case if $w\lesssim 50~\MeV$. 

If the example hidden sector that we consider has a portal coupling $\lambda \gtrsim 10^{-7}$ at a scale $w\gtrsim 10~\GeV$, it reaches thermal equilibrium with the visible sector and is cosmologically safe. $\phi$ decays safely before BBN and the relic abundance of hidden sector gauge bosons can easily be viable. Such a model is also not excluded by collider bounds, e.g. on invisible Higgs decays, provided that $\lambda$ is not too large. 

It is hard, but not impossible, to accommodate the hidden sector that we consider if $50~\MeV \lesssim w\lesssim 5~\GeV$ and it is at the same temperature as the visible sector. $\phi$ must decay to evade the relic density constraint, however light hidden sector fermions cannot be introduced due to BBN limits. Instead $\phi$ must decay to the visible sector before BBN, which most easily happens through a Higgs portal operator. There are numerous strong constraints on such operators, and such a model is only viable in small parts of parameter space (e.g. if $\phi$ has a mixing angle with the Higgs $\sin^2 \theta \simeq 10^{-8}$) \cite{Flacke:2016szy}. 

More generally, we expect that there are relatively few viable models of hidden sectors that are thermalised with the visible sector and which go through transitions at temperatures $10~\MeV \lesssim T \lesssim 5~\GeV$. Such a sector must have a relatively strong interaction with the visible sector for its energy to be transferred to the visible sector before BBN. However, portal interactions at low scales are strongly constrained by, for example, observations of supernovae and beam dump experiments.

\section{Phase transitions} \label{sec:phaset}

A first order phase transition begins when bubbles of the true vacuum start to be nucleated at a significant rate and completes when these have expanded and engulfed the universe. In the process the difference in energy density between the two vacua, including the finite temperature contribution to the potential, is released. This drives the expansion of the bubbles and heats the thermal bath behind the bubble walls.\footnote{In transitions with subsonic walls speeds, the plasma ahead of the bubble walls is also heated.} 

Bubbles of the true vacuum can be nucleated by thermal fluctuations or by quantum tunnelling through the energy barrier. Given the scaling of the volume and surface energies of a bubble with its radius, there is a minimum bubble size above which it will expand. The probability of nucleating such a bubble is set by an action, determined from the field profile of a critical bubble. The action depends on whether the bubble is nucleated through a thermal fluctuation or quantum tunnelling and is denoted $S_3$ or $S_4$ respectively.

The probability of nucleating a critical bubble via a thermal fluctuation per unit time and volume, $\Gamma_3$, is approximately
\beq \label{eq:GammaS3}
\Gamma_3 \simeq T_{\rm h}^4 \left(\frac{S_3}{2\pi T_{\rm h}} \right)^{3/2} e^{-S_3/T_{\rm h}} ~,
\eeq
where $T_{\rm h}$ is the temperature of the hidden sector \cite{Linde:1981zj}. The analogous expression for the rate of nucleation by tunnelling, $\Gamma_4$, is 
\begin{equation}\label{eq:GammaS4}
\Gamma_4 \simeq w^4 \left(\frac{S_4}{2\pi}\right)^2 e^{-S_4} ~,
\end{equation}
where $w$ is the VEV that $\phi$ gets after the transition \cite{Affleck:1980ac,Turner:1992tz}. As well as the explicit temperature dependence in $S_3/T_{\rm h}$, both $S_3$ and $S_4$ implicitly depend on the hidden sector temperature, since they are determined by the finite temperature potential. Both $S_4$ and $S_3/T_{\rm h}$ tend to infinity for temperatures approaching that at which the high temperature phase is energetically favoured and go to $0$ if the barrier disappears.

The critical actions can sometimes be estimated with one of several analytic approximations \cite{Anderson:1991zb}. The simplest of these, appropriate if the barrier between the vacua is large compared to the energy difference between them, is the thin wall approximation. This treats the actions as a sum of independent contributions from the bubble's volume and from its surface. However, the assumption of a thin wall is not valid for the close to conformal models that we consider and leads to significant inaccuracies \cite{Jaeckel:2016jlh} (other analytic approaches are also inaccurate in such models \cite{Huang:2018fum}).\footnote{Calculating the predicted thickness of the bubble walls in our model assuming the thin wall approximation shows that this is not self consistent, and the results for the critical actions that we obtain numerically also deviate significantly from the thin wall predictions.}

Instead, we calculate the critical actions numerically by varying the field configurations to minimise the actions.\footnote{This can be done efficiently using an overshoot/undershoot method. We use the publicly available code CosmoTransitions \cite{Wainwright:2011kj}, which we have validated with our own implementation.} 
We also note that Eqs.~\eqref{eq:GammaS3} and \eqref{eq:GammaS4} are only approximations to the nucleation rates. Their accuracy can be improved by replacing the factors in front of the exponentials with a function that includes the determinant of fluctuations around the critical field configuration \cite{Strumia:1998qq,Strumia:1999fv,Strumia:1998nf,Strumia:1998vd}. However, the error introduced by Eqs.~\eqref{eq:GammaS3} and \eqref{eq:GammaS4} is typically relatively mild, so they are sufficient for our present purposes.\footnote{Alternative, more accurate, non-perturbative methods to calculate the nucleation probabilities have also been developed \cite{Moore:2000jw}.}

A transition is guaranteed to complete if the hidden sector potential is such that the barrier between the two minima disappears at low temperatures. Provided that the true vacuum is energetically favoured before the barrier vanishes (and internal thermal equilibrium is maintained), the transition happens through bubble nucleation before the barrier vanishes completely due to the rapid increase in the $\Gamma_3$ and $\Gamma_4$ at such times \cite{Mintz:2008rt,Bessa:2008nw}. The temperatures at which the symmetry breaking vacuum is energetically favoured, that at which nucleation becomes efficient, and that at which the barrier disappears completely are typically not dramatically different.

In parts of the parameter space in which a barrier remains between the two vacua at zero temperature, $S_3/T_{\rm h}$ initially decreases as the hidden sector temperature is decreased before reaching a minimum at some finite temperature. As the temperature drops further $S_3/T_{\rm h}$ increases, since the energy available from the thermal bath to fluctuate into a bubble decreases. Meanwhile, because the barrier approaches a constant shape in the $T_{\rm h}\rightarrow 0$ limit, $S_4$ asymptotes to a non-zero value. 

\begin{figure}[t]\begin{center}
\includegraphics[height=5.2cm]{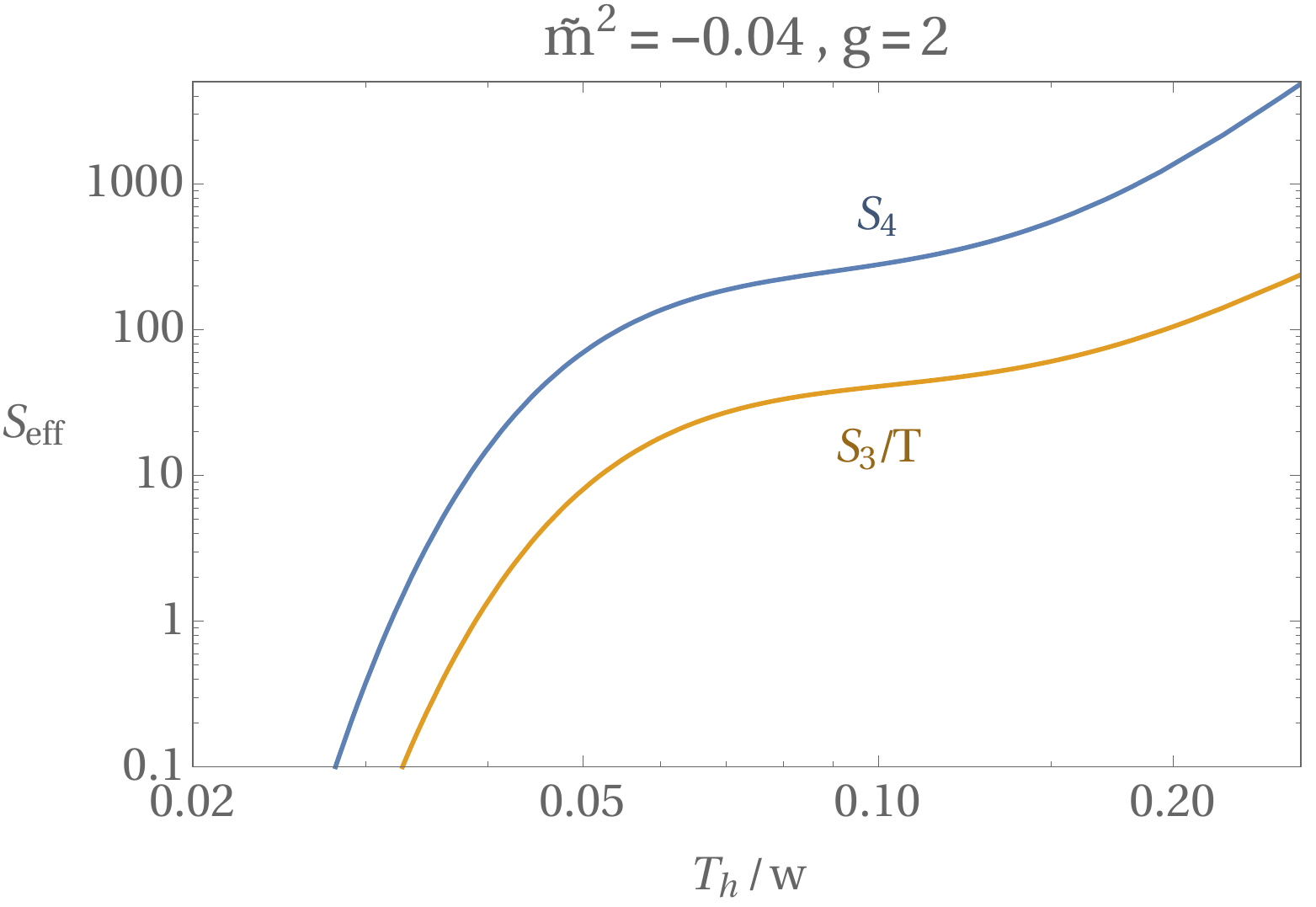}\hspace{0.05cm}
\includegraphics[height=5.1cm]{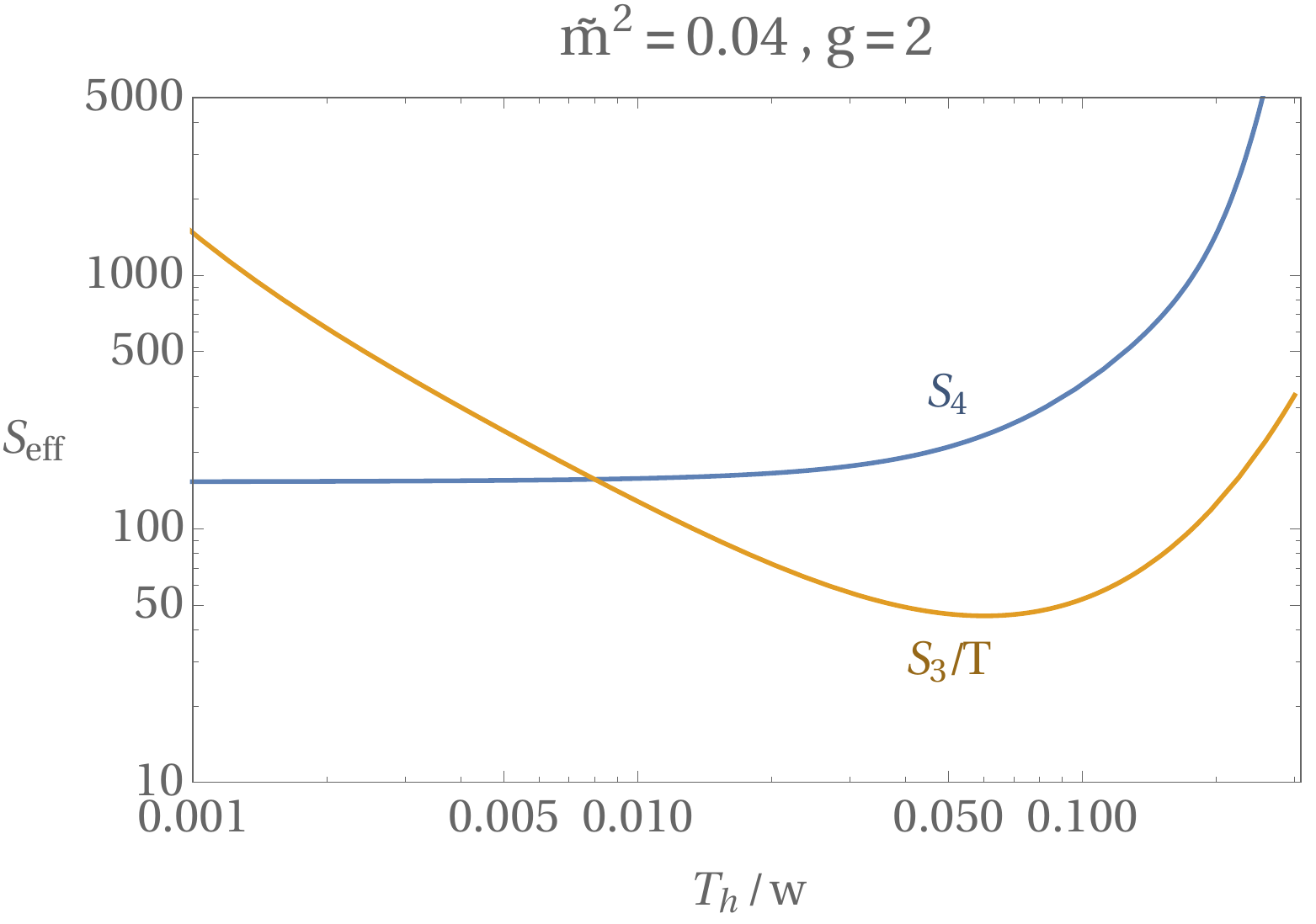} \end{center}
\caption{Examples of the dependence of the critical bubble nucleation actions $S_3/T_{\rm h}$ and $S_4$ on the hidden sector temperature $T_{\rm h}$. The parameters $g$ and $\tilde{m}^2$ define the model via Eq.~\eqref{eq:v0}. The left panel corresponds to a hidden sector for which the energy barrier between the two minima vanishes at zero temperature, so both $S_3/T_{\rm h}$ and $S_4$ go to zero. The right panel corresponds to a model in which a barrier remains at zero temperature, so $S_4$ asymptotes to a constant and $S_3/T_{\rm  h} \rightarrow \infty$.
}
\label{fig:act}
\end{figure}

In Figure~\ref{fig:act} we show examples of the critical bubble actions for two points in the parameter space of the hidden sector model described in Section~\ref{sec:model}.

The time when a phase transition begins can be estimated as the point when one critical bubble is nucleated per Hubble volume per Hubble time, so that the true vacuum starts to permeate, i.e. when
\beq \label{eq:simpcond}
{\rm max}\left[\Gamma_3\left(T_{\rm h}\right),\Gamma_4\left(T_{\rm h}\right) \right] = H\left(T_{\rm v}\right)^4 ~.
\eeq
If a hidden sector is at the same temperature as the visible sector, this condition is satisfied if $S_3/ T_{\rm h} \simeq 120$ or $S_4 \simeq 120$ for a phase transition at temperatures around the EW scale.

Although Eq.~\eqref{eq:simpcond} is useful to get a rough idea of when a transition happens, it is not accurate enough to reliably determine whether a transition successfully completes if a barrier remains at zero temperature. Further, the gravitational wave signal that is produced depends on how long a transition takes and the average bubble size. To extract these properties, we evaluate $S_3/T_{\rm h}$ and $S_4$ as a function of temperature for a given point in hidden sector parameter space. Another required physical input is the speed of the bubble walls throughout the transition $v_{\rm w}$. Over all of the parameter space that we consider the phase transitions are fairly strong so that $v_{\rm w} \simeq 1$ and for tracking the progress of the transition and the average size of bubbles it is sufficient to fix $v = 1$.\footnote{When we study the gravitational wave signals produced, the difference between bubble walls with Lorentz factors $\gamma_{\rm w} = \mathcal{O}\left(1\right)$ and those with $\gamma_{\rm w} \rightarrow \infty$ will be important.} 
Then we track the proportion of the universe that is in the low temperature phase and the distribution of bubble sizes throughout the phase transitions, accounting for bubbles only forming in regions of space that are in the false vacuum and allowing for the expansion of bubbles. 
This calculation is standard (see e.g. Section 3.2 of \cite{Megevand:2016lpr} for a clear summary).

\subsection{Transitions with hot and cold hidden sectors}\label{subsec:Thc}

We start by assuming that the hidden sector reheating temperature is high enough that the high temperature phase  ($ \left<\phi\right> =0$) is initially favoured. As before the temperature of the visible and hidden sectors are allowed to differ by a ratio $\epsilon$, and we assume there is no energy transfer between the sectors.\footnote{When presenting results we quote $\epsilon$ at the time of the hidden sector phase transition.}

\begin{figure}[t]\begin{center}
\includegraphics[height=7cm]{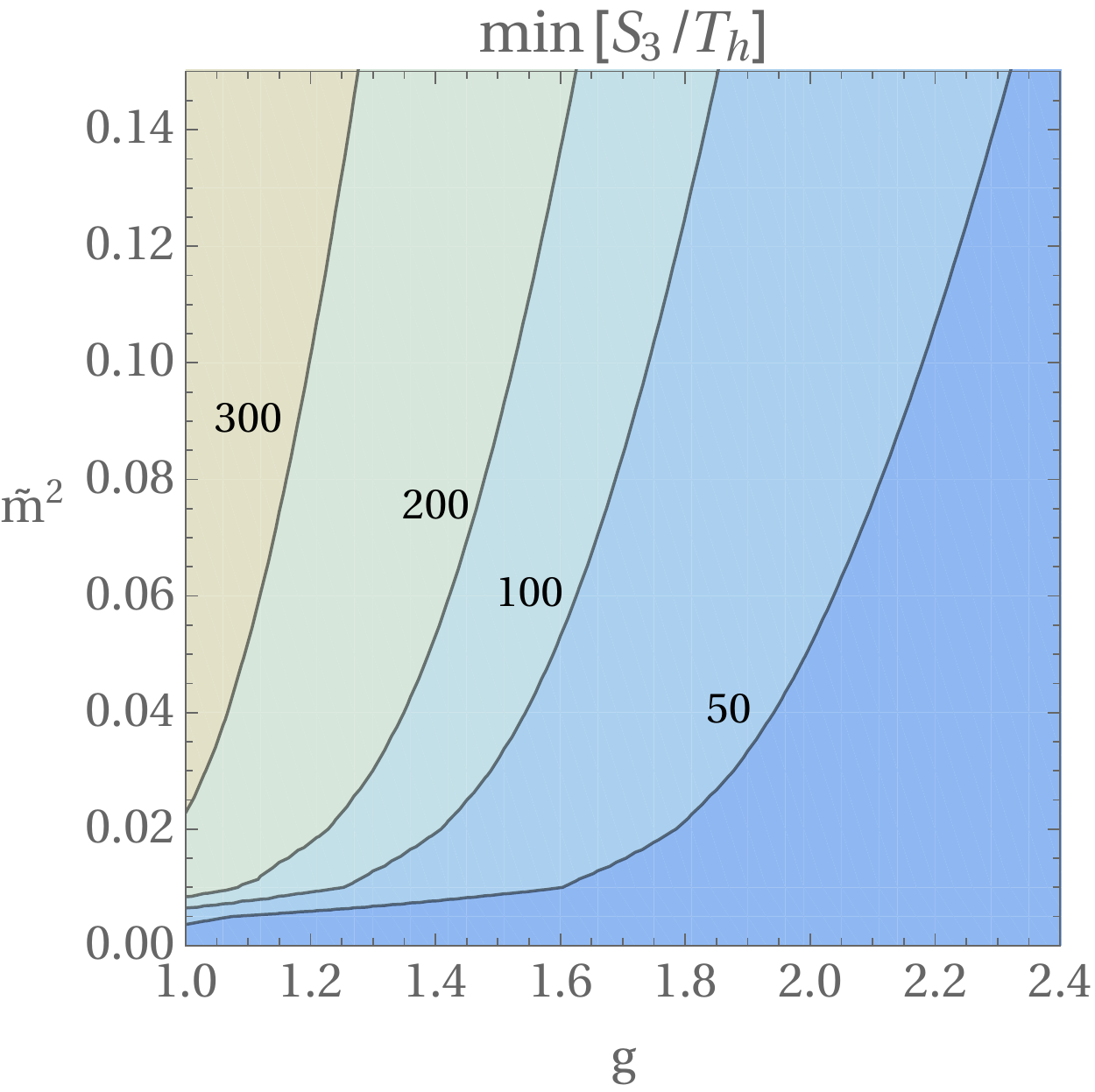}\hspace{0.5cm}
\includegraphics[height=7cm]{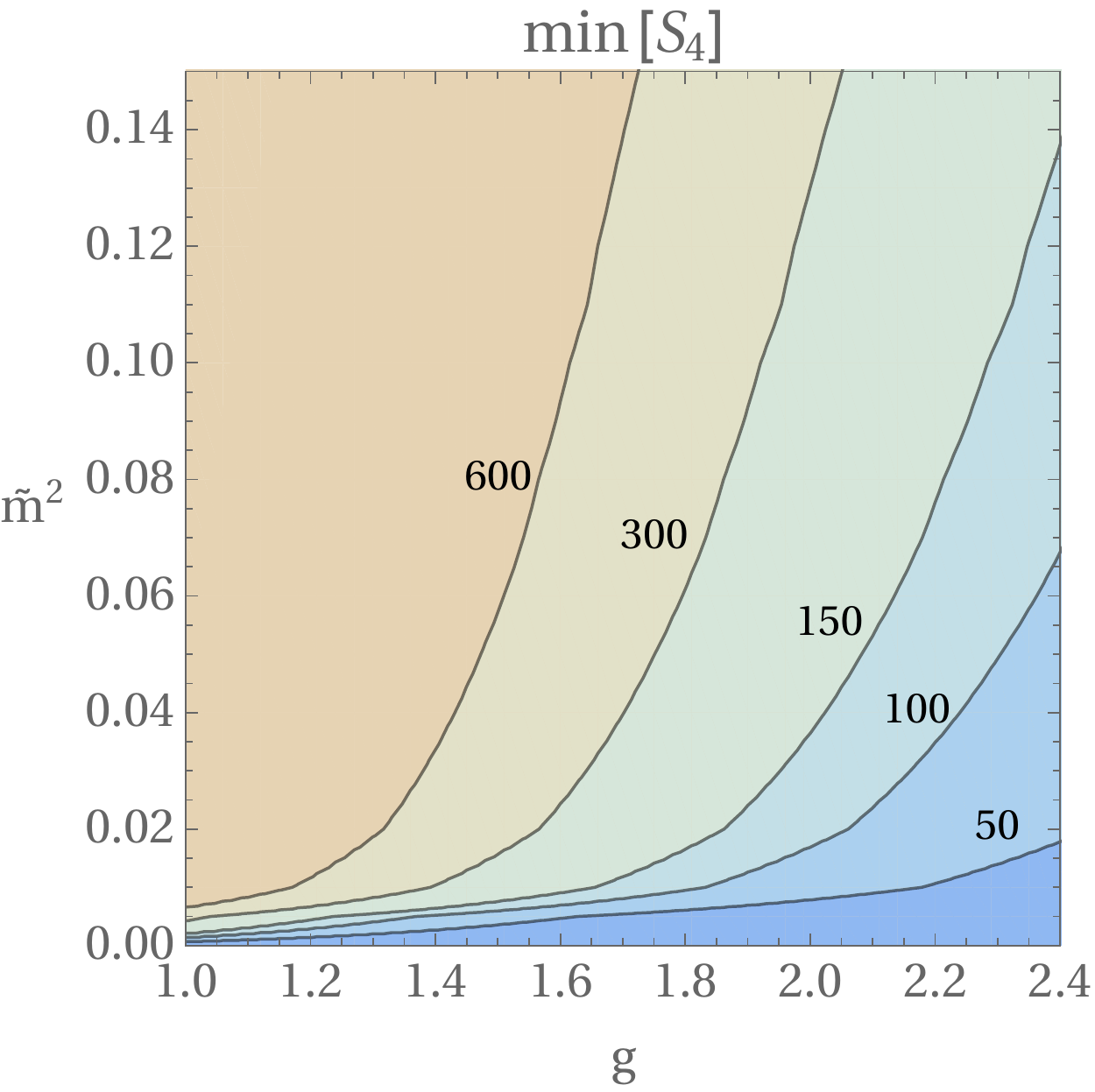} \end{center}
\caption{Contours of the minimum values of $S_3/T_{\rm h}$ (left) and $S_4$ (right) as a function of the parameters of the hidden sector (defined in Section~\ref{sec:model}).}
\label{fig:contours}
\end{figure}

In Figure~\ref{fig:contours} we plot contours of the minimum values of $S_4$ and $S_3/T_{\rm h}$ as a function of the dimensionless parameters of the hidden sector, for models such that a barrier persists at zero temperature. The value of $w$ does not affect these results, since it is the only relevant scale in the calculation. We consider relatively large gauge couplings $1\gtrsim g \gtrsim 2.5$. Towards the upper end of this range the accuracy of our perturbative calculations may be compromised, however since $g<\sqrt{4\pi}$ we do not expect the qualitative dynamics to change significantly. Therefore, despite this source of potential numerical imprecision, we regard our hidden sector as a useful toy model to explore the phenomenological possibilities that can arise more generally.

The minimum value of $S_3/T_{\rm h}$  is smaller than that of $S_4$ over all of the parameter space in Figure~\ref{fig:contours}. Further, $S_3/T_{\rm h}$ is also smaller than $S_4$ at hidden sector temperatures $T_{\rm h} \sim w$ (as is the case for the points in parameter space shown in Figure~\ref{fig:act}). These features are not surprising. As discussed in \cite{Espinosa:2008kw}, in the thin wall approximation the actions scale as
\beq \label{eq:s3scal}
S_{3}/T_{\rm h} \sim \frac{w}{T_{\rm h}} \left(\frac{w^4}{\Delta V}\right)^2 ~,
\eeq
and
\beq \label{eq:s4scal}
S_4 \sim \left(\frac{w^4}{\Delta V} \right)^3 ~,
\eeq
where $\Delta V$ is the difference in energy density between the two vacua. 
Even though the thin wall approximation often does not give precise numerical results, the feature that if $S_3/T_{\rm h} \gg 1$ then $S_4 \gg S_3/T_{\rm h}$ is typical across many classes of models (although it would be interesting to find theories for which it does not hold). 
As a result, if a first order phase transition happens at a temperature $T_{\rm h} \sim w$, this will be through nucleation of bubbles by thermal fluctuations (including in the case that no barrier remains at zero temperature).

In a model for which a barrier remains at zero temperature, if a transition does not complete when $T_{\rm h} \sim w$ then  $S_3/T_{\rm h}$ subsequently increases as the temperature drops further, while $S_4$ only changes slightly. This raises the possibility that a hidden sector might fail to complete a thermal transition but could later undergo a tunnelling transition once the Hubble parameter has dropped, despite $\Gamma_4$ remaining smaller than the largest value of $\Gamma_3$. However, if the hidden and visible sectors are at the same temperature this is not possible in generic models, for a reason related to the problems faced by old inflation \cite{Guth:1982pn}. If a transition is to occur, the total vacuum energy density at the true minimum must be tuned to (approximately) zero,  so the vacuum energy density of the false minimum is $\sim w^4$. Once $T_{\rm h}=T_{\rm v}\ll w$ this will dominate the energy density of the universe. As a result $H$ remains $\sim  w^{2}/M_{\rm Pl}$, and the universe enters a new inflationary phase. Therefore, the proposed tunnelling transition, which requires much smaller values of $H$ since $S_4 \gg \rm {min}\left(S_3/T_{\rm h}\right)$, cannot occur.

Consequently, for the model we consider, a first order phase transition in a hidden sector at the same temperature as the visible sector will only ever happen through nucleation of bubbles by thermal fluctuations, at a time when the hidden sector temperature is $T_{\rm h}\sim w$. We also believe that this is a typical feature across generic models, although it would be interesting to study other calculable models further.

This conclusion does not hold if the hidden sector is cold relative to the visible sector. If there is barrier between the two vacua that remains at zero temperature, the Hubble parameter when $S_3/T_{\rm h}$ is minimised is $H \sim \epsilon^{-2} w^2/M_{\rm Pl}$. For a transition to occur through thermal fluctuations requires
\beq
H\left(T_{\rm v}\right)^4 \lesssim T_{\rm h}^4 \left(\frac{S_3}{2 \pi T_{\rm h}}\right)^{3/2} e^{-S_3/T_{\rm h}} ~.
\eeq
If this is ever satisfied, it happens when $T_{\rm h}\sim w$, and requires
\beq \label{eq:s3epcond}
\left(\frac{S_3}{T_{\rm h}}\right) < \left(\frac{S_3}{T_{\rm h}}\right)_{\rm 0} - 8 \log\left(\frac{1}{\epsilon}\right) ~,
\eeq
where $\left(S_3/T_{\rm h}\right)_{\rm 0}$ is the value necessary for a transition to complete when the two sectors are at equal temperatures (e.g. $\left(S_3/T_{\rm h}\right)_{\rm 0} \sim 120$ for an EW scale transition). A cold hidden sector requires a smaller value of $S_3/T_{\rm h}$ because the universe is expanding faster when $S_3/T_{\rm h}$ reaches its minimum, so the condition for the true vacuum to permeate is stronger.

On the other hand, the vacuum energy of a cold hidden sector still starts dominating the expansion of the universe when $H_{\rm min} \simeq \sqrt{\rho_{\rm vac}}/\left(3 M_{\rm Pl}\right) $, and provided 
\beq \label{eq:s4epcond}
\Gamma_4\left(T_{\rm h}\simeq 0\right) \gtrsim H_{\rm min}^4 ~,
\eeq
the transition can complete through tunnelling. This is independent of whether the hidden sector is colder than the visible sector (apart from the variation of $\Gamma_4$ with temperature, which is negligibly small for the relevant temperatures $T_{\rm h}\ll w$).

\begin{figure}[t]\begin{center}
\includegraphics[height=4.8cm]{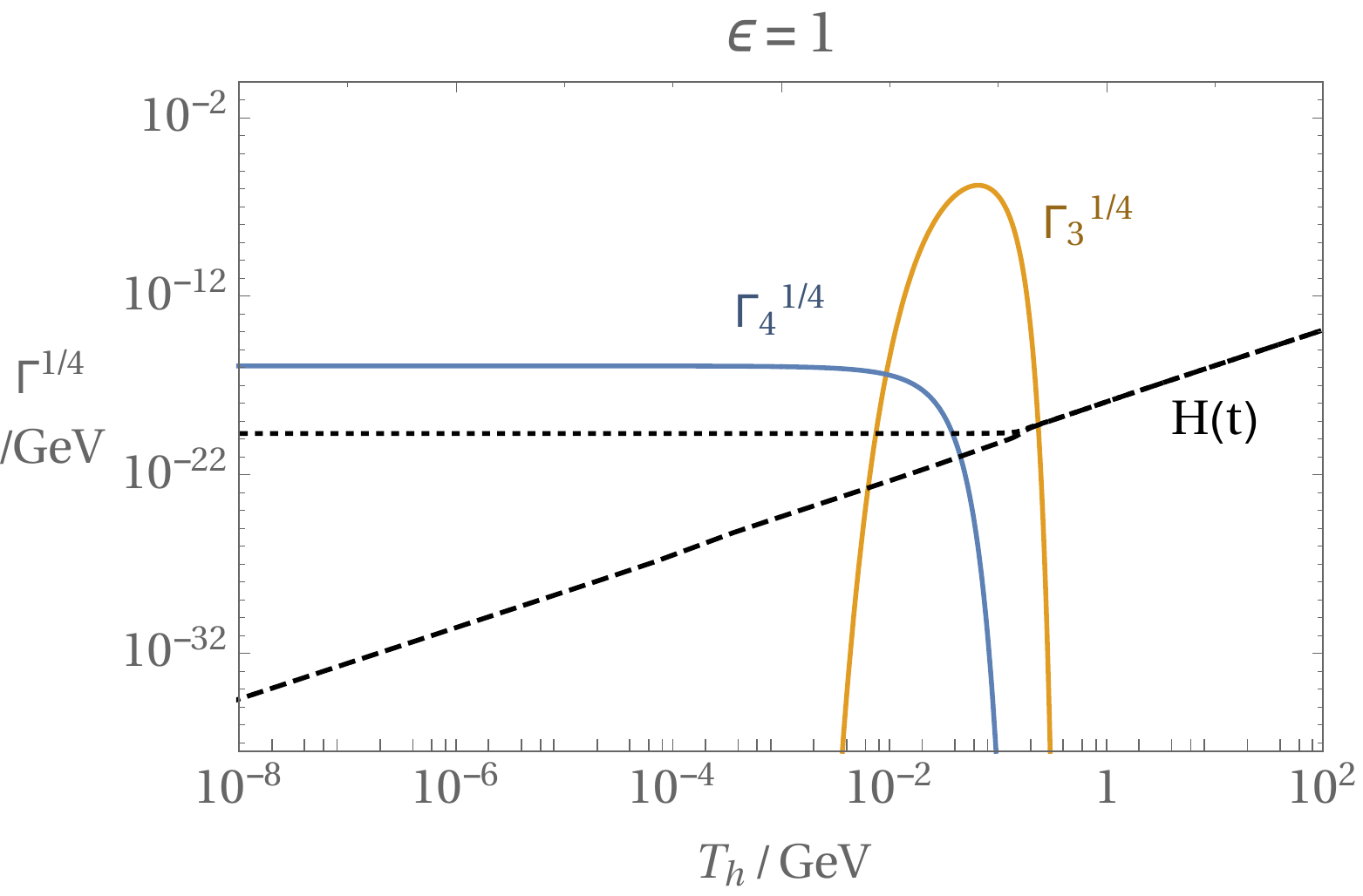}\hspace{0.20cm}
\includegraphics[height=4.8cm]{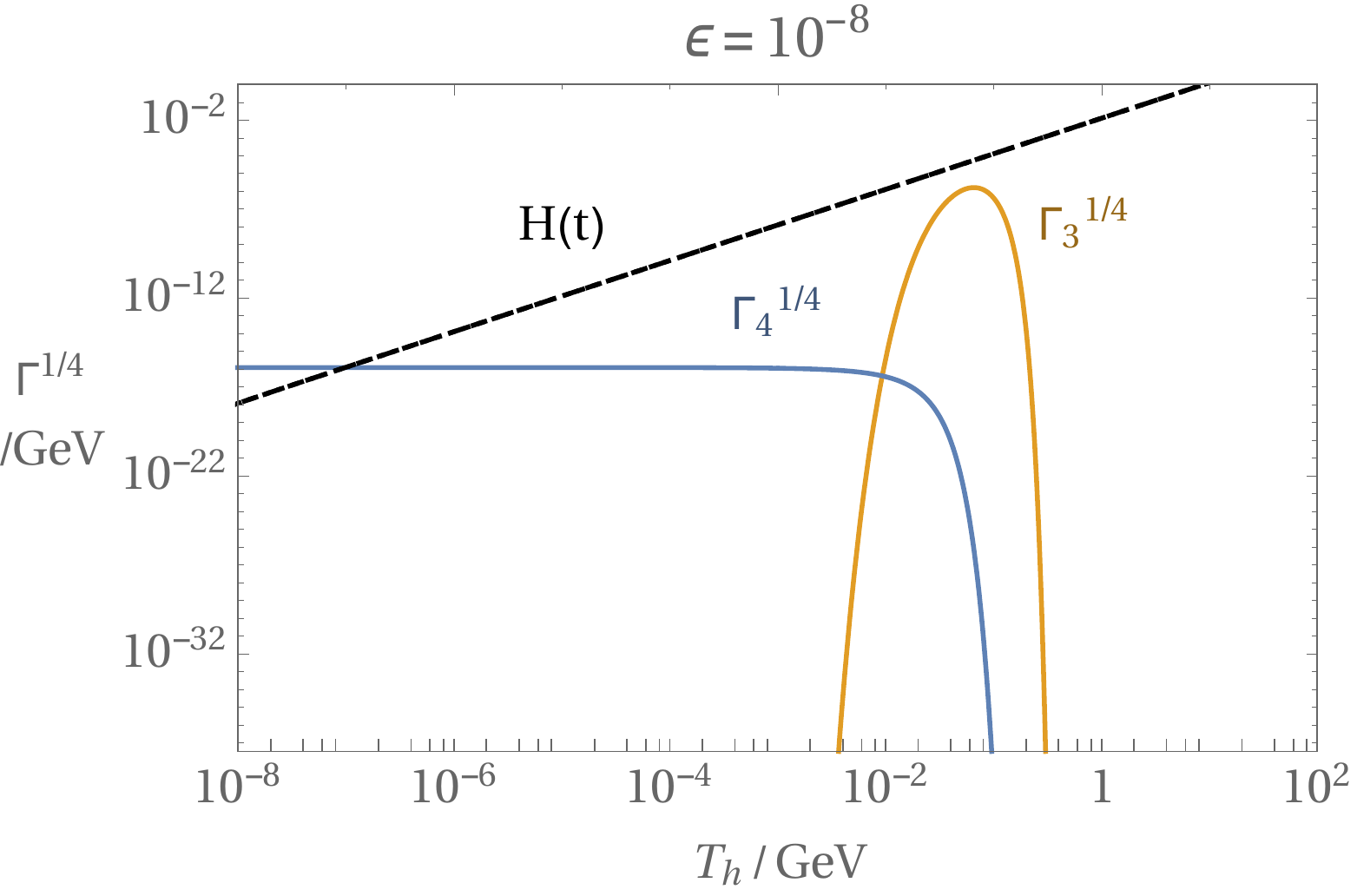} \end{center}
\caption{The bubble nucleation rate as a function of the hidden sector temperature for a model with $g=2$ and $\tilde{m}^2=0.04$ (as in the left panel of Figure~\ref{fig:act}) and $w=1~\GeV$. Results are shown for a hidden sector at the same temperature as the visible sector ($\epsilon=1$, left) and for a hidden sector that is much colder ($\epsilon=10^{-8}$, right). The Hubble parameter in the two cases is also plotted, assuming that the transition occurs prior to the hidden sector false vacuum energy dominating the energy density of the universe (dashed black), and assuming the phase transition does not complete prior to this (dotted black). If $\epsilon=1$ (left panel) the hidden sector false vacuum energy begins to dominate at $T_{\rm v}= T_{\rm h} \simeq 0.2 \thinspace \rm \GeV$ when the two Hubble curves diverge. A transition happens when $H(t)$ is first smaller than one of $\Gamma_3^{1/4}$ or $\Gamma_4^{1/4}$. Therefore the warm hidden sector (left) undergoes a thermally nucleated transition, while the cold hidden sector (right) misses a thermal transition, but subsequently goes through a tunnelling transition.}
\label{fig:gammaH}
\end{figure}

Eq.~\eqref{eq:s4epcond} is a weaker condition than Eq.~\eqref{eq:s3epcond}, so for sufficiently small $\epsilon$ a hidden sector can fail to undergo a thermal transition at $T_{\rm h}\sim w$, but later goes through a tunnelling transition. Even though the minimum value of $S_3/T_{\rm h}$ is smaller than that of $S_4$, the tunnelling transition happens later when the visible sector temperature has dropped and the Hubble parameter is smaller. 
To demonstrate this we plot the nucleation rates via thermal fluctuations and tunnelling as a function of the hidden sector temperature in Figure~\ref{fig:gammaH}, for a model with $\epsilon =1$ and $\epsilon=10^{-8}$. The hidden sector is the same as in the left panel of Figure~\ref{fig:act} with $w=1~\GeV$. We also plot the Hubble parameter assuming that the transition occurs prior to the hidden sector false vacuum energy density dominating the universe and assuming that the transition never occurs, with the cosmological constant tuned to zero in the true vacuum. However, in both models plotted the transition will complete and the dotted Hubble dependence is not realised. As expected, for $\epsilon =1$ the transition happens through thermal nucleation.\footnote{If the hidden sector parameters are altered to increase the height of the barrier, $\Gamma_3$ and $\Gamma_4$ decrease by approximately the same factor. So if the maximum value of $\Gamma_3$ is small enough that a thermal transition does not occur, the hidden sector false vacuum energy density dominates the universe before a tunnelling transition is possible}. For $\epsilon=10^{-8}$, the larger value of the Hubble parameter when $T_{\rm h} \sim w$ prevents a thermal transition occurring, and a tunnelling transition happens later once $H$ is smaller.

For a tunnelling transition to happen this way, $\log \epsilon$ must be comparable to the difference between the minimum values of $S_3/T_{\rm h}$ and $S_4$. Since this is typically $\mathcal{O}(10)$ a huge temperature hierarchy is required, indeed, in the model that we consider $\epsilon \lesssim 10^{-7}$. Such small values are compatible with our assumed cosmological history, provided that the visible sector reheating temperature $T_{\rm RH} \gtrsim w/\epsilon$ (so that the hidden sector is reheated above $w$). The visible sector temperature when the phase transition takes place is fixed by 
\beq
\Gamma_4 \simeq H\left(T_{\rm v}\right)^4 ~,
\eeq
i.e. when
\beq \label{eq:TvisTun}
T_{\rm v} \simeq \frac{S_4^{1/4} e^{-S_4/8}}{g_v^{1/4}}  \sqrt{w M_{\rm Pl}} ~, 
\eeq
where $S_4$ is approximately temperature independent at the relevant times. The exponential dependence on $S_4$ in Eq.~\eqref{eq:TvisTun} means that $T_{\rm v}$ changes by orders of magnitude as the parameters of the hidden sector vary.

\begin{figure}[t]\begin{center}
\includegraphics[height=7cm]{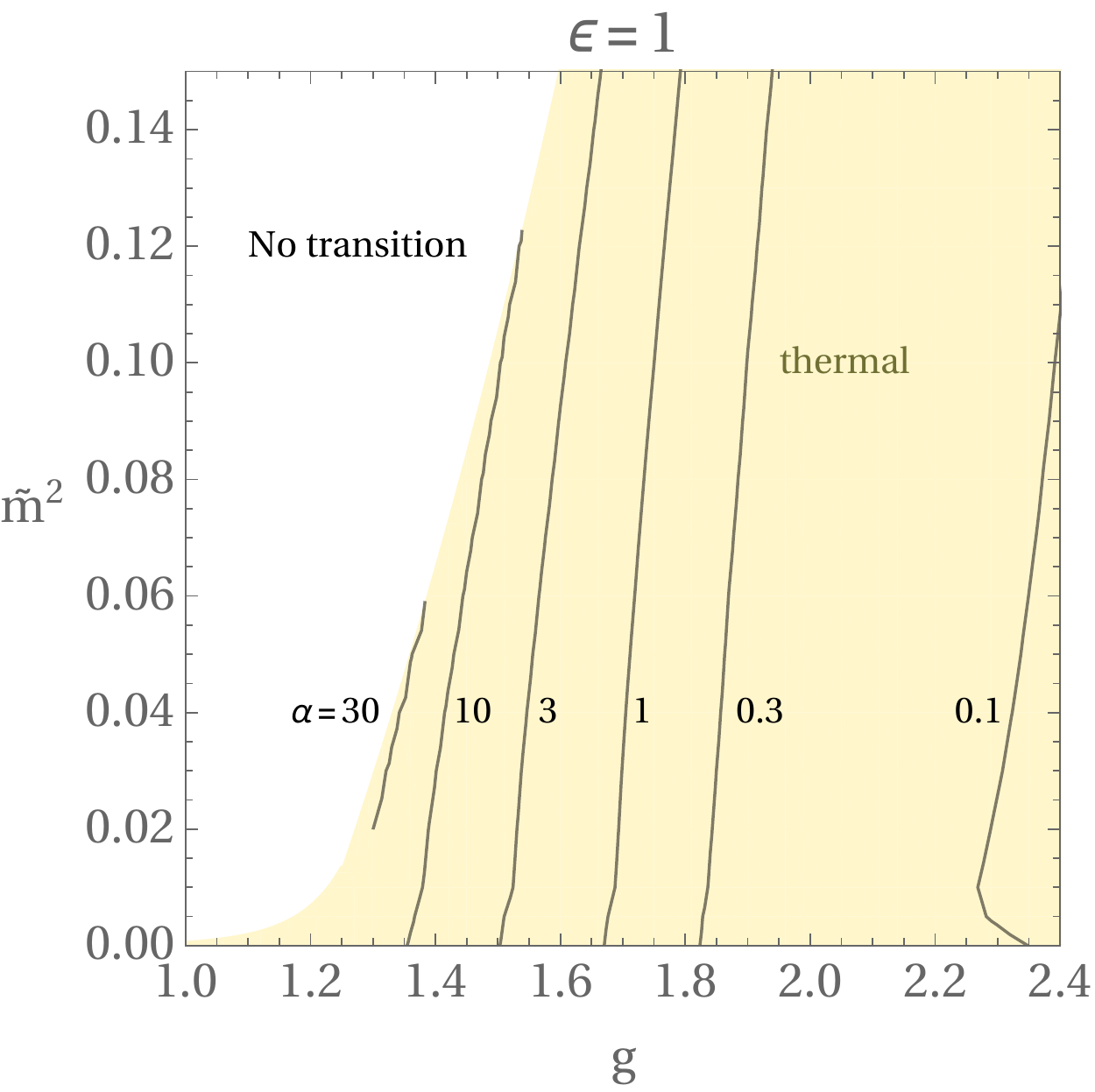}\hspace{0.25cm}
\includegraphics[height=7cm]{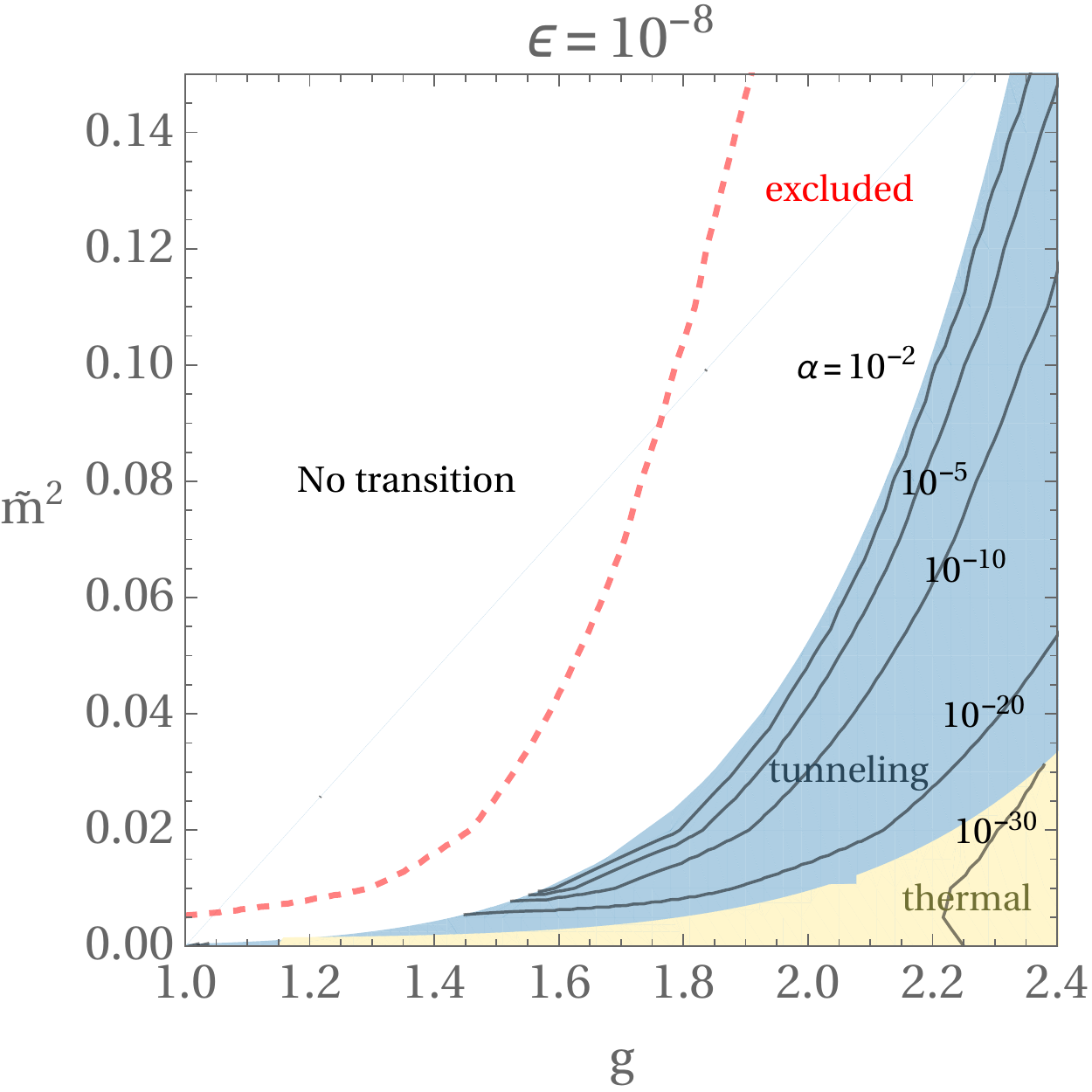} \end{center}
\caption{The type of phase transition that occurs (yellow: thermal, blue: tunnelling) over the hidden sector parameter space for $w=1~\GeV$, when the hidden sector is at the same temperature as the visible sector (left) and when the hidden sector is much colder with $\epsilon =10^{-8}$ (right). The contours give the values $\alpha$ for the transitions. In the white region no phase transition takes place. For cold hidden sectors there are models that cannot lead to an acceptable cosmological history, as described in the text.}
\label{fig:typeofT}
\end{figure}

In Figure~\ref{fig:typeofT} we plot the type of phase transition that occurs over the parameter space of our example hidden sector for $\epsilon =1$ and $\epsilon =10^{-8}$. Only parameter space in which a barrier remains at zero temperature is shown, since in the converse case a thermal transition will always occur.

Some models in Figure~\ref{fig:typeofT} right are incompatible with the cosmological history of the universe regardless of how the vacuum energy is tuned, despite the hidden sector being extremely cold. For example, both tuning the vacuum energy of the universe to zero when the hidden sector is in the false vacuum and tuning it to zero in the true minimum lead to an unacceptable cosmology when the thermal transition window is missed. In the former case, the tunneling nucleation rate must be small compared to the present day value of the Hubble parameter, which is not always the case, whilst in the latter case the universe can become dominated by the false minimum and subsequently trapped in this phase. A significant region of the parameter space of models with $\epsilon =10^{-8}$ fail in both scenarios and are always problematic.\footnote{If $\epsilon =1$ the difference between $S_3/T_{\rm h}$ and $S_4$ is sufficiently large that if a thermal transition is missed, a tunnelling transition will be slow compared to the age of the universe, so this issue does not arise.}

Finally we note that models with a low reheating temperature, below that at which the hidden sector temperature is restored, do not evade our argument that tunnelling transitions only occur in cold hidden sectors. We do not consider such theories further, and details of the dynamics in this case may be found in Appendix~\ref{app:lowreheat}.

\subsection{Properties of the phase transitions}

If a hidden sector is colder than the visible sector this will affect the properties of its phase transition, even if the transition still happens through thermal fluctuations; and if a transition happens through tunnelling rather than thermal fluctuations this will also affect its dynamics. 

One crucial quantity for determining the gravitational wave signal produced is the amount of energy released by the phase transition, relative to that in the thermal bath. This is the quantity $\alpha$ defined in Eq.~\eqref{eq:alpha}.

First we consider transitions that happen through thermal fluctuations. Suppose there are two identical hidden sectors, at the same scale $w$, that both go through thermal transitions, one of which is at the same temperature as the visible sector and the other much colder. Both transitions happen when their respective hidden sector temperatures are $T_{\rm h}\sim w$, with only an $\mathcal{O}(1)$ difference due to the increased Hubble parameter when there is a hotter visible sector present (c.f. Eq.~\eqref{eq:s3epcond}). Therefore the visible sector temperature at the time of the phase transition is approximately
\beq
T_{\rm v} \propto \frac{w}{\epsilon} ~,
\eeq
and
\beq \label{eq:alphaepi}
\alpha =  \frac{g_{\rm h}}{g_{\rm v}} \alpha_{\rm h} \epsilon^4 ~.
\eeq
In thermal transitions the relative energy released into the hidden sector is $\alpha_{\rm h} \lesssim 10$ and $\alpha$ is strongly suppressed when the hidden sector is cold. 

The situation is different if a hidden sector transition happens through tunnelling. In this case the visible sector temperature at the time of the transition, and therefore $\alpha$, is fixed by  Eq.~\eqref{eq:TvisTun}, which is independent of $\epsilon$. Unlike for thermal transitions, $\alpha$ can span a wide range of values for different models. It still cannot be orders of magnitude larger than $1$ since this would mean that the hidden sector vacuum energy dominated the universe prior to the transition, leading back to the problems of old inflation. Since tunnelling transitions always happen at temperatures $T_{\rm h}<w$ we also know that $\alpha \gtrsim \epsilon^4$.

Contours of $\alpha$ as a function of the hidden sector parameters are plotted in Figure~\ref{fig:typeofT}. In the left panel $\epsilon =1$, so the phase transition is always via thermal fluctuations and $\alpha$ is roughly $\mathcal{O}(1)$. As the value of $g$ increases the hidden sector potential favours the true vacuum at higher temperatures, so the transition happens slightly earlier and $\alpha$ decreases. 
Relatively large values of $\alpha \simeq 30$ are possible close to the boundary at which the transition only just manages to complete, corresponding to significant supercooling.  This is due to the almost conformal nature of our hidden sector, which results in the energy barrier between the two vacua remaining temperature dependent down to temperatures significantly below $w$. For example, Figure~\ref{fig:pot} shows the barrier changing at temperatures around $T \simeq 0.1 w$.\footnote{In that figure $\tilde{m}^2$ is relatively large so that the presence (or absence) of the barrier is visible. Models with smaller values of $\tilde{m}^2$ are more phenomenologically interesting, and in this case the suppression of the critical temperature is more pronounced.} The possibility that conformal models could lead to significant supercooling has previously been studied in \cite{Konstandin:2010cd,Konstandin:2011dr,Huang:2018fum,Brdar:2018num}.
 
In the right panel of Figure~\ref{fig:typeofT} the hidden sector is cold, with $\epsilon =10^{-8}$, and over the majority of this parameter space a phase transition happens through tunnelling. As expected, $\alpha$ takes a wide range of values $ 10^{-25}$ -- $1$ and it is fairly large only close to the boundary at which a transition fails to complete. In the part of parameter space for which a thermal transition takes place $\alpha$ is extremely small, roughly $\epsilon^{4}$.

Another parameter that is important in determining the gravitational wave signal is the time taken for the phase transition to complete relative to the Hubble parameter when the transition occurs. This is also approximately inversely proportional to the average size of bubbles when they collide compared to the Hubble distance.

If a thermal transition takes place relatively fast, and at a temperature such that $S_3/T_{\rm h}$ decreases approximately linearly with temperature, it will complete once $S_3$ decreases by an order 1 amount after nucleation first becomes efficient (since its exponential dependence means the nucleation rate will be extremely fast at this point). The duration of such a phase transition can be estimated as $\beta^{-1}$ where
\beq \label{eq:bdef1}
\beta = - \frac{\partial S_3}{\partial t} ~,
\eeq
evaluated at the time of the phase transition (see e.g. \cite{Caprini:2015zlo}). 
Eq.~\eqref{eq:bdef1} can be rewritten in the more useful form
\beq 
\begin{aligned} 
\frac{\beta}{H_*} &= T_{{\rm v}*} \frac{\partial}{\partial T_{\rm v}} \left( \frac{S_3}{T_{\rm h}}\right)\\
&= T_{{\rm h}*} \frac{\partial}{\partial T_{\rm h}} \left( \frac{S_3}{T_{\rm h}}\right)~,
\label{eq:beta}
\end{aligned}
\eeq
where a $*$ indicates a quantity at the time of the phase transition, and the derivative is evaluated at this time as well. The right hand side of Eq.~\eqref{eq:beta} is not far from $\mathcal{O}\left(1\right)$ even if the hidden sector is cold, so the duration of such a phase transition is always approximately set by the Hubble parameter.

More generally the duration of a transition can be defined as the time between e.g. $90\%$ and $10\%$ of the universe being in the high temperature phase (the qualitative features of our results are not sensitive to these particular choices). Unlike the estimate from $\beta$, this is applicable to tunnelling transitions, and also thermal transitions that happen when $S_3/T_{\rm h}$ is close to its minimum (additionally, Eq.~\eqref{eq:beta} requires modification in the case of significant supercooling, as can occur in our model). 

In tunnelling transitions the duration of a phase transition is also approximately set by the Hubble parameter. Once nucleation is efficient enough that $10\%$ of the universe reaches the low temperature phase, existing bubbles expand at close to the speed of light and further bubbles continue to form, so the rest of space goes through the transition within approximately a Hubble time. This is also the case in the previously mentioned classes of thermal transitions for which $\beta$ is not a good measure of the duration.

\begin{figure}[t]\begin{center}
\includegraphics[height=7cm]{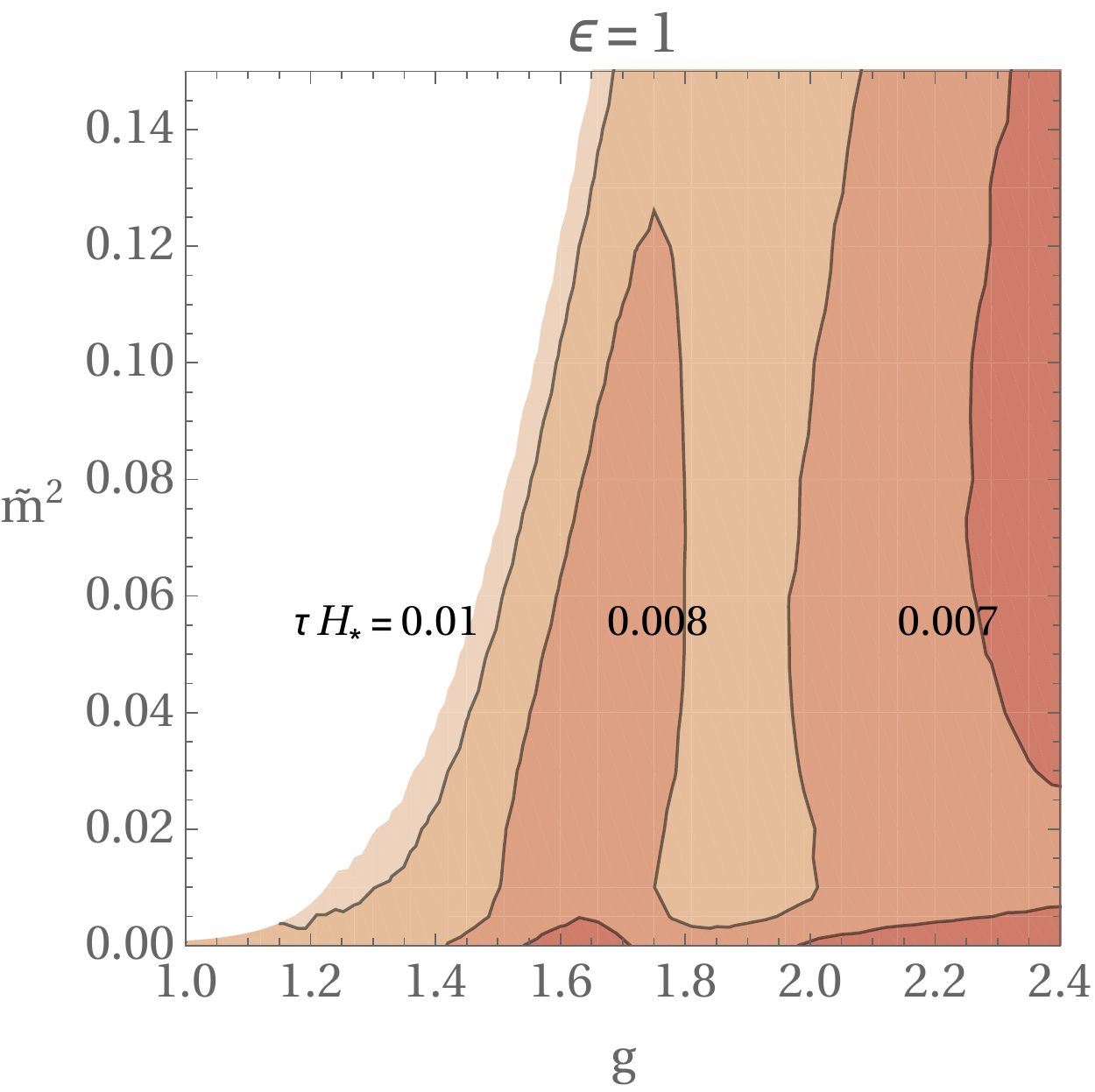}\hspace{0.5cm}
\includegraphics[height=7cm]{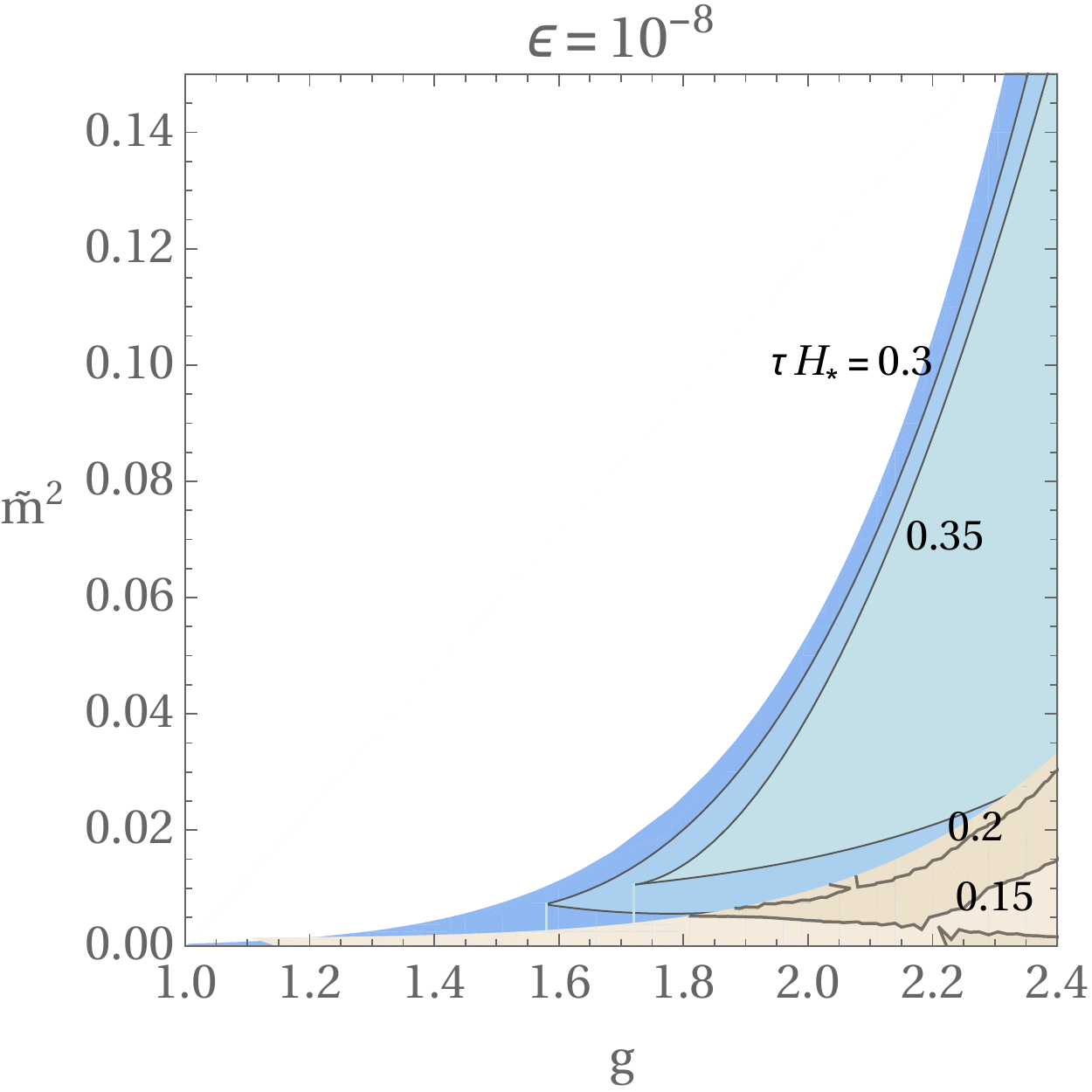}
\end{center}
\caption{Contours of the duration of the phase transition $\tau$ relative to the Hubble parameter when it occurs, for a hidden sector at the same temperature as the visible sector (left) in which case the transition happens via thermal fluctuations (red shading), and for a cold hidden sector with $\epsilon =10^{-8}$ (right), in which case the bubbles are nucleated through tunnelling over most of the parameter space (blue shading).}
\label{fig:time}
\end{figure}

We plot the duration of the hidden sector phase transition as a function of the model's parameters  in Figure~\ref{fig:time}, for $\epsilon =1$ and $10^{-8}$. As expected the duration of a transition is $\sim H^{-1}_*$ up to a numerical factor $\lesssim 100$, regardless of whether the hidden sector is cold. The average bubble radius can also be calculated and is similarly parametrically given by $H^{-1}_*$, even if the hidden sector is cold (with a numerical factor $\sim 0.001$ -- $0.1$, as expected from \cite{Hogan:1984hx}).

Despite having the same parametric dependence on $H_*$, there is a mild difference between the duration of a thermal and tunnelling transitions. In typical thermal transitions the critical action decreases fast once it first becomes small enough for a significant number of bubbles to form, and the transition usually completes within $\simeq 1/100$ of a Hubble time.\footnote{The small non-monotonic dependence on $g$ that is visible in Figure~\ref{fig:time} left is because exact dependence of $S_3/T_{\rm h}$ on time throughout the phase transition varies across the parameter space (e.g. transition might begin closer or further away from the minimum of $S_3/T_{\rm h}$).}
Tunnelling transitions typically take slightly longer, leading to a slightly larger average bubble radius, since the nucleation rate is constant in this case.\footnote{For $\epsilon =10^{-8}$ thermal transitions are slightly slower than when $\epsilon$ is larger. This is because such transitions only just complete, and happens when $S_3/T_{\rm h}$ is close to its minimum and approximately temperature independent.}

\section{Bubble wall velocities in hot and cold transitions} \label{sec:wallv}

We now study the velocities of bubble walls in models with thermal and tunnelling transitions. The speed that bubble walls reach is related to the proportion of the energy released by the phase transition that is concentrated in the bubble walls compared to the proportion being dissipated into the surrounding plasma. 
As we discuss in the next section, this has a significant effect on the spectrum of gravitational waves that is produced by a phase transition.

\subsection{Friction on bubble walls}\label{subsec:friction}

If a transition happens when the hidden sector temperature is non-zero, the bubble walls  pass through a bath of hidden sector particles as they expand. The interactions of these states with the bubble walls transfer part of the energy released by the phase transition to the plasma.

In all of the parameter space that we consider the phase transition is relatively strong, so the bubble walls expand faster than the speed of sound. This is a typical feature of theories that produce observable gravitational wave signals, since a fairly strong transition is required for significant gravitational wave emission. Consequently, the number density of hidden sector states $n_{\rm h}$ in front of the bubble wall is determined by the temperature of the hidden sector immediately prior to the transition (the situation is more complicated in subsonic transitions due to a shock front).

The dynamics of bubble walls can be described in terms of a driving pressure, sourced by the difference in energy densities between the meta-stable and true vacua, and a frictional pressure $P_{\rm fr}$ acting against the expansion of the bubble walls. The velocity of a given bubble wall, $v_{\rm w}$, changes with time as
\beq \label{eq:dvdtb}
\frac{d v_{\rm w}}{d t}= \frac{1}{\sigma \gamma_{\rm w}^3} \left(\rho_{\rm vac} - P_{\rm fr} \right) ~,
\eeq
where $\gamma_{\rm w}$ is the relativistic Lorentz factor of the bubble walls, and $\sigma$ is their energy per unit area (i.e. their tension).

At leading order in the gauge coupling (in an expansion in $g^2/\left(4\pi\right)$) the friction on the bubble walls is due to states having different masses in the two phases. This means  there is a mismatch between the thermal distribution of states crossing the bubble walls and the equilibrium configuration in the true vacuum, and evolving to the equilibrium distribution transfers energy into the plasma.

The friction that is produced by this effect is independent of $\gamma_{\rm w}$ when $\gamma_{\rm w} \gg 1$. In this limit the total effective pressure on the bubble walls can be obtained by making the replacement
\beq \label{eq:pfr1}
\left( \rho_{\rm vac}-P_{\rm fr} \right) \rightarrow \left( V_0\left(0\right)-V_0\left(\phi_{\rm min}\right)+ \frac{T_{\rm h}^2}{24} \sum_i m_i^2\left(0\right)-m_i^2\left(\phi_{\rm min}\right)\right) ~,
\eeq
in Eq.~\eqref{eq:dvdtb}, where $\phi_{\min}$ is the position of the true vacuum of the full finite temperature potential, and $V_0$ is the zero temperature potential.\footnote{Eq.~\eqref{eq:pfr1} is only an approximation to the friction from this source. More accurate expressions in terms of integrals over particle occupation numbers are given in \cite{Bodeker:2009qy}.} In our example model the right hand side of Eq.~\eqref{eq:pfr1} is
\beq 
V_0\left(0\right)-V_0\left(\phi_{\rm min}\right)+ \frac{T_{\rm h}^2}{24} \sum_i m_i^2\left(0\right)-m_i^2\left(\phi_{\rm min}\right) = \frac{9g^4}{1024 \pi^2}\frac{2-\tilde{m}^2}{4}w^4 - \frac{3 g^2 w^2 T_{\rm h}^2 }{32} ~,
\eeq
where we have consistently neglected the subleading contribution from $\phi$ itself, as in Section~\ref{sec:model}.

If, in the limit $\gamma_{\rm w} \rightarrow \infty$, the friction from this source is greater than the driving force then the right hand side of Eq.~\eqref{eq:pfr1} is positive, and the bubble walls reach a finite terminal velocity. As observed in \cite{Bodeker:2009qy} this is the case if the true vacuum is not favoured in the mean field approximation to the thermal potential. On the other hand, if the friction is smaller than the driving force this source of friction will not prevent the bubble walls accelerating indefinitely. 

Not surprisingly, the friction and driving forces in Eq.~\eqref{eq:pfr1} depend only on the temperature of the hidden sector and its microscopic properties. Therefore, if a hidden sector goes through a thermal transition, the friction on the bubble walls is approximately independent of whether it is cold relative to the visible sector. There are only order 1 changes in the friction due to bubble nucleation becoming efficient at slightly different hidden sector temperatures, as a result of the larger value of the Hubble parameter at this time if the hidden sector is cold. 
We find that this source of friction leads to finite bubble wall speeds over some, but not all, of the parameter space of the model that we consider. The region with finite bubble walls speeds approximately coincides with those parts of  Figure~\ref{fig:typeofT} for which $\alpha \lesssim 0.5$, and the value of $\epsilon$ only changes the boundary of this region of parameter space slightly.\footnote{Other conditions for this source of friction to prevent bubble walls accelerating without bound have been given in the literature, sometimes differing slightly from Eq.~\eqref{eq:pfr1}. The small differences are of no consequence for our present work.}

If there is not enough friction in the plasma to impede the bubble walls expansion they are said to `runaway' where they accelerate without bound right up until the point they collide with each other. Although our discussion so far would seem to permit runaway bubble walls in thermal transitions in some parts of parameter space, there can be other sources of friction that do not lead to a constant pressure in the limit $\gamma_{\rm w} \rightarrow \infty$. In particular, if a gauge boson changes mass across the wall, as is the case in the model we study, splitting radiation leads to a friction term that grows $\propto \gamma_{\rm w}$ \cite{Bodeker:2017cim}. Consequently, the bubble walls have a maximum velocity if there is a bath of hidden sector particles present.  

There is still uncertainty on the exact parametric dependence of the friction from splitting radiation. \cite{Bodeker:2017cim} proposes that it scales as
\beq
F_{\rm fr} \sim \gamma_{\rm w} g^3 w T_{\rm h}^3 ~,
\eeq
although they also discuss a possible weaker dependence on $\gamma_{\rm w}$. The exact scaling of this friction deserves further study, and will affect out quantitative determination of the boundaries between regimes, but it does not affect the qualitative possibilities that we identify. Similarly to the $\gamma_{\rm w}$ independent friction, this source of friction is only dependent on the hidden sector temperature. Therefore in thermal transitions, the value of the bubble wall's terminal velocity is independent of of whether the hidden sector is cold relative to the visible sector. Since it is not too far from $\mathcal{O}\left(1\right)$, a bubble wall quickly reaches this terminal value shortly after it is nucleated.

It is currently unknown whether sectors in which no gauge boson masses change across the phase transition can have runaway bubbles, or if there is another source of friction that scales as $\gamma_{\rm w}^n$ with $n>0$. Given that $\gamma_{\rm w}$ reaches extremely large values in runaway transitions, $\sim 10^{14}$, such a contribution could prevent runaway walls even if it has an extremely suppressed coefficient, e.g. from multiple loop factors.\footnote{A $\propto \log \gamma_{\rm w}$ dependence might not be enough to stop the bubble walls accelerating before the transition completes.} This is a major source of uncertainty on the dynamics of the bubble walls in such models, and it is an important topic to resolve in the future.

\subsection{Runaway vs non-runaway walls}\label{subsec:runawayvsnon}

In thermal transitions with finite bubble wall speeds, the terminal values of $\gamma_{\rm w}$ are not too far from $\mathcal{O}\left(1\right)$ regardless of which source of friction dominates.\footnote{This is why we do not study the boundary between these two regimes in detail.} The bubble walls reach such speeds very quickly after nucleating, long before they typically collide. The energy density in bubble walls is $\simeq \gamma_{\rm w} \sigma R_*^2$,  where $R_*$ is the average bubble size at collision, and the energy released by the region of space now inside the bubble is $\sim \rho_{\rm vac} R^3$. Therefore, since the average bubble size when they collide is $\sim H^{-1}_* \sim (\epsilon^2M_{\rm Pl})/w^2$, the energy in bubble walls is negligible compared to that in the plasma when the bubbles percolate.\footnote{This assumes that $\epsilon$ is not tiny. If $\epsilon$ is extremely small a thermal transition is less likely, and even if one occurs the gravitational wave signal will be unobservable.}

Even if hidden sectors without gauge bosons can have thermal transitions with runaway bubbles, part of the energy released by such phase transitions will be transferred to the plasma in this case, through the leading order friction Eq.~\eqref{eq:pfr1}. For an explicit model, assuming that only the $\gamma_{\rm w}$ independent friction is present, the proportion of the released energy transferred to the plasma can be found by considering the hydrodynamic solutions of the bubble wall \cite{Espinosa:2010hh}. Apart from extremely strong transitions, with significant supercooling, at least $\sim 10\%$ of the energy is goes into the plasma, with the remainder localised in the bubble walls.

We now argue that, in contrast, tunnelling transitions can lead to the vast majority of the energy density going into bubble walls with a negligible proportion transferred to the plasma via friction. This is possible because the hidden sector can be arbitrarily cold compared to the scale that sets the driving force in such models. 

The hidden sector temperature at the time of a tunnelling transition is $T_{\rm h}= \epsilon T_{\rm v} \simeq (\epsilon g w)/\alpha^{1/4}$. Therefore the $\gamma_{\rm w}$ independent contribution to the friction Eq.~\eqref{eq:pfr1} is suppressed by $T_{\rm h}^2 \simeq \epsilon^2 g^2/ \alpha^{1/2}$. For small $\epsilon$, and not too small $\alpha$, this does not transfer a significant fraction of the released energy to the plasma, and it is never sufficient to prevent the bubble walls running away. This is intuitively due to the low hidden sector temperature compared to the scale of the driving force $T_{\rm h}\ll w$ suppressing the number density of the hidden sector particles that the walls pass through, and also reducing the mismatch in the thermal distributions ahead of and behind the bubble walls.

The $\gamma_{\rm w}$ dependent contribution to the friction has a different effect to the $\gamma_{\rm w}$ independent piece, since it grows arbitrarily large as the bubble walls gain speed. However, for a sufficiently cold hidden sector the bubble walls will not have reached sufficiently high speeds for this friction to become relevant before they collide. If they are still accelerating at the time of the collision, the friction will be suppressed by $ \gamma_*/\gamma_{\rm t}$ where $\gamma_*$ is the Lorentz factor of the bubble walls when they collide and $\gamma_{\rm t}$ is the value corresponding to the bubble wall's terminal velocity. Apart from models close to the boundary $\gamma_* = \gamma_{\rm t}$, such a suppression is sufficient to prevent any significant energy transfer to the plasma. On the other hand, if the bubble walls reach their terminal velocity before colliding, at least an order 1 fraction of the released energy goes into the plasma, and if they reach terminal speeds long before they collide the vast majority of the energy goes into the plasma. 

In the model that we consider the $\gamma_{\rm w}$ dependent frictional force is parametrically $\ n_{\rm h} \gamma_{\rm w} g^3 w $, and the driving force on the bubble walls per unit area is approximately $g^4 w^4$ (ignoring numerical factors). The terminal value of $\gamma_{\rm w}$ is therefore roughly

\beq
\gamma_{\rm t} \simeq \frac{w^3}{n_{\rm h}} \simeq \frac{w^3}{\epsilon^3 T_{{\rm v}*}^3} \simeq \frac{\alpha^{3/4} }{g^2 \epsilon^3} ~.
\eeq
As expected, the terminal velocity is large if $\epsilon$ is small and $\alpha$ is not too small.

We compare $\gamma_{\rm t}$ to what the wall's Lorentz factor would be in the absence of friction as a function of a bubble's radius, which we denote $\gamma_0\left(R\right)$. In the thin wall approximation, which is accurate up to order 1 factors in our model, this is simply
\beq
\gamma_{0}\left(R\right) \simeq \frac{R \rho_{\rm v}}{\sigma} \simeq g^2 R w ~.
\eeq
The typical bubble radius at the time of collisions $R_*$ is to be set by the Hubble parameter at the time of transition
\beq
R_{*} \simeq 10^{-2} H^{-1}_* \simeq \alpha^{1/2} \frac{M_{\rm Pl}}{g^2 w^2} .
\eeq
As a result, the bubble walls are still accelerating at the time they collide if
\beq
\alpha^{1/2} \frac{M_{\rm Pl}}{w} \lesssim \frac{\alpha^{3/4}}{g^2  \epsilon^3} ~,
\eeq
that is
\beq \label{eq:epraw}
\epsilon   < 10^{-6} \alpha^{1/12} \left(\frac{w}{{\rm GeV}}\right)^{1/3} g^{-2/3}~.
\eeq
Thus a hidden sector must be extremely cold for the bubble walls to effectively runaway.

For a particular hidden sector, the energy density in the bubble walls when they collide compared to if there was no friction is
\beq \label{eq:gammara}
\frac{\gamma_*}{\gamma_0}={\rm min}\left[1, R_{\rm t}/R_{*} \right] = {\rm min}\left[1,  \frac{\alpha^{1/4} }{g^2 \epsilon^3}  \frac{w }{ M_{\rm Pl} }  \right] ~,
\eeq
where $\gamma_*/\gamma_0=1$ corresponds to runaway bubble walls. The proportion of the released energy density that goes into the fluid is, to a good approximation, $1- \gamma_*/\gamma_0$, which is negligible for runaway bubbles and $\simeq 1$ for bubbles that reach their terminal velocity.


\section{Gravitational waves}\label{sec:gvsignals}

We now consider the gravitational wave spectra produced by first order transitions, in hot and cold hidden sectors. Rather than focusing exclusively on the particular hidden sector that we study, we aim to study the effect of those features identified above that apply to large classes of hidden sectors.

Gravitational waves from phase transitions can be produced by different processes, and excellent reviews on the development of the literature can be found in  \cite{Binetruy:2012ze,Weir:2017wfa,Caprini:2009fx,Caprini:2018mtu,Croon:2018new} . Depending on the type of transition, and the dynamics of the bubble walls, the overall spectrum is made up of contributions from a subset of the following sources:

\begin{itemize}
    \item \textbf{Colliding scalar field shells}. Depending on the nature of the phase transition, a significant fraction of the energy released can be concentrated in the bubble walls. When these collide they lead to quadrupole moments, which efficiently emit gravitational waves.
    
    \item \textbf{Acoustic waves in the plasma}. If a transition happens in a thermal bath, some of the energy in the wall will be deposited into the plasma via friction. This produces acoustic wave fronts in the plasma which, when they collide, can source gravitational waves. 
    
    \item \textbf{Turbulence in the plasma}. After the acoustic sound shells  collide, some portion of their energy is transferred into turbulent flows, which could act as a relatively long lasting source of gravitational waves.

    \item \textbf{Long lived field oscillations after collisions}. When bubble collision take place they can establish long lived oscillations of the field, which can emit gravitational waves. 

\end{itemize}

The gravitational wave spectrum emitted by each of these sources can be studied using numerical simulations and theoretical models. Although both of these approaches has shortcomings, a reasonably good understanding of the expected signal from each source has been reached in the literature, as a function of the physical parameters of a transition.\footnote{In this Section, we denote the transition time by $\beta$, regardless of the type of transition (even though we previously defined $\beta$ in Eq.~\eqref{eq:beta} in a way that was only appropriate to particular classes of thermal transitions).} For our present work, gravitational waves from bubble collisions and sound waves are the most important, while turbulence and long lived field oscillations give negligible contributions to the signal. We therefore focus on these, utilising standard parameterised fits to predict the signals produced. In Appendix~\ref{app:GW}, we collect results from different parts of the literature that support this approach.

\subsection{Bubble collisions}

The gravitational wave signal from colliding bubbles during a first order phase transition was first proposed in \cite{Witten:1984rs,1986MNRAS.218..629H}  and was carefully studied in \cite{Kosowsky:1991ua,Kosowsky:1992rz,Kosowsky:1992vn}, where the `envelope' approximation was developed.  This approximates the gravitational waves from collisions as being sourced by an expanding infinitely thin bubble of stress energy, and neglects regions in which the bubbles have previously
overlapped. In this context the thickness of a bubble wall is judged relative to e.g. bulk motions of any plasma present, and the thin wall assumption was shown to be valid for relatively strong transitions in \cite{Kamionkowski:1993fg}.\footnote{This is a weaker condition than that involved in the thin walled approximation for calculating the critical bubble actions.} In cold hidden sectors the profile of the bubble walls is determined by $\alpha_{\rm h}$ rather than $\alpha$, and the thin wall assumption is valid for all the models that we consider.

Various predictions, both analytical \cite{Kosowsky:1992rz,Jinno:2016vai,Jinno:2017ixd} and numerical \cite{Huber:2008hg,Weir:2016tov,Cutting:2018tjt}, have been made regarding the form of the frequency dependence of the gravitational wave spectrum emitted by bubble collisions. 
We use the fit of the gravitational wave power spectrum from bubble collisions given in \cite{Weir:2017wfa}. After redshifting to the present day, this takes the form
\begin{equation}
\Omega_\text{coll}(f)\equiv \frac{1}{\rho_\text{crit}}\frac{d\rho_\text{GW-coll}}{d\ln f}, 
\end{equation}
where $\rho_\text{crit}$ is the critical density of the Universe, and
\begin{equation}
  \label{eq:scalaransatz}
  h^2 \Omega_\text{coll}(f) = 1.67 \times 10^{-5} \, \Delta
  \left(\frac{H_*}{\beta}\right)^2 \left( \frac{\kappa_\phi \alpha}{1 +
    \alpha} \right)^2 \left(\frac{100}{g_{{\rm v*}}} \right)^{\frac{1}{3}}
  S_\text{env}(f)
\end{equation}
where $h\simeq 0.7$ is the dimensionless present day Hubble parameter. As before, $*$ indicates the value of a quantity at the time of the phase transition, and $\alpha$ is the ratio between the energy density released during the transition and the background radiation density in the visible sector defined in Eq.~\eqref{eq:alpha}.

The remaining parameters in Eq.~\eqref{eq:scalaransatz} are: $\kappa_{\phi}$, which is the efficiency with which the vacuum energy released is deposited in the bubble wall; $\Delta$, which is the amplitude of the gravitational wave signal in the limit $\kappa_\phi \rightarrow 1$ and $\alpha\gg 1$; and $S_\text{env}(f)$, which is the spectral shape, normalised to have maximum value 1.

$\Delta$ can be fitted by
\beq
\Delta =
 \frac{0.48  v_\mathrm{w}^3}{1 + 5.3 v_\mathrm{w}^2 + 5
  v_\mathrm{w}^4}~ .
\eeq

Using the envelope approximation, the peak energy density in gravitational waves from thin walled bubble collisions scales like $h^2\Omega_{env} \propto \kappa_\phi^2$.  The dependence of the peak energy density on $\kappa_{\phi}$ in Eq.~\eqref{eq:scalaransatz} can be derived in the envelope approximation, and is also supported by numerical simulation. 
The overall amplitude, i.e. the prefactor in Eq.~\eqref{eq:scalaransatz}, is set by theory \cite{Jinno:2016vai} and agrees fairly well with simulations \cite{Huber:2008hg}. 
We can approximate 
\beq \label{eq:kphidef}
\kappa_\phi \simeq \frac{\gamma_{\rm w}\sigma}{R_*\rho_{vac}}~,
\eeq
where $\sigma$ is the surface tension of the wall and $R_*$ is the average bubble separation length at collision, $R_*=v_\mathrm{w}\beta^{-1}$. This is precisely the quantity that we calculated when analysing the finite bubble wall speeds, leading to the result Eq.~\eqref{eq:gammara}.

Finally, the frequency dependence of the gravitational wave spectrum $S_\text{env}(f)$ takes the form (for $v_\mathrm{w}$ close to 1)
\begin{equation}
  S_\text{env}(f) = \left[ c_l
    \left(\frac{f}{f_\text{env}}\right)^{-3} + (1 - c_l -
    c_h)\left(\frac{f}{f_\text{env}} \right)^{-1} + c_h
    \left(\frac{f}{f_\text{env}}\right) \right]^{-1}
\end{equation}
where fits to numerical simulations give rise to the values $c_l = 0.064$ and $c_h = 0.48$. We are assuming here that the high frequency tail drops like $f^{-1}$, which as discussed may not be precisely the case. The peak frequency $f_*$ is given by
\begin{equation}
  \label{eq:peakfreqenv}
f_\text{env} = 16.5  \, \mu\mathrm{Hz} \,
\left(\frac{f_*}{\beta} \right) \left( \frac{\beta}{H_*} \right)
\left( \frac{T_{v*}}{100 \, \mathrm{GeV}} \right) \left( \frac{g_*}{100}
\right)^{\frac{1}{6}} ~,
\end{equation}
which has a dependence on $v_\mathrm{w}$ via \footnote{This has a slightly different form to that adopted in \cite{Breitbach:2018ddu} due to the fact they are using results from the earlier analysis of \cite{Huber:2008hg}.}
\begin{equation}
\frac{f_*}{\beta} = \frac{0.35}{1 + 0.069
  v_\mathrm{w} + 0.69 v_\mathrm{w}^4}.
\end{equation}

\subsection{Sound waves}

If a significant proportion of the energy released by a phase transition is transferred to the plasma through friction, sound waves in the plasma form. These propagate through the primordial plasma either behind the wall, as deflagrations, or in front of it as detonations (hybrid regimes are also possible). As mentioned, the model that we consider is always in the supersonic detonation regime for both thermal and tunnelling transitions. The collision of these acoustic shells causes a stirring of the plasma, which provides a long lasting source of gravitational waves.

A combination of numerical simulations \cite{Hindmarsh:2013xza,Hindmarsh:2015qta,Hindmarsh:2017gnf} and analytical models \cite{Hindmarsh:2016lnk,Jinno:2017fby} suggest that sound waves are an important source of gravitational waves if a significant fluid component exists. The spectrum obtained peaks at a wavelength approximately set by the average bubble separation at collision, $R_*$, and at high frequencies, the signal falls off $\propto f^{-3}$ for detonations, and it seems to be even steeper for deflagrations. This is in stark contrast to the shape of signals arising from phase transitions occurring in vacuum, which fall off in the range $f^{-1}$ to $f^{-1.5}$ at high frequencies.

For our present work, we use the fit of the gravitational wave spectrum given in \cite{Weir:2017wfa}, based on the simulations in  \cite{Hindmarsh:2017gnf}. This is
\begin{equation} \label{eq:swparam}
h^2 \Omega_\text{sw}(f) =
8.5 \times 10^{-6} \left(\frac{100}{g_*} \right)^{\frac{1}{3}}\kappa_\mathrm{sw}^2(v_w, \alpha) \alpha^2 \left(\frac{H_*}{\beta}\right)
v_\mathrm{w} \, S_\text{sw}(f) ~,
\end{equation}
where the 
spectral shape is
\begin{equation} \label{eq:Ssw}
   S_\text{sw}(f) = \left(\frac{f}{f_\mathrm{sw}}\right)^3  \left(
   \frac{7}{4 + 3 (f/f_\mathrm{sw})^2 } \right)^{7/2} ~,
\end{equation}
with approximate peak frequency 
\begin{equation} \label{eq:peaksound}
  f_\mathrm{sw} = 8.9 \, \mu\mathrm{Hz} \, 
  \frac{1}{v_\mathrm{w}} \left( \frac{\beta}{H_*} \right) \left( \frac{T_{v*}}{100 \,
    \mathrm{GeV}}\right) \left( \frac{g_*}{100} \right)^\frac{1}{6} ~,
\end{equation}
and we have fixed the simulation derived factor $z_\mathrm{p}$ which appears in \cite{Hindmarsh:2017gnf} to take what is estimated to be its usual value of $10$.

The fit in Eq.~\eqref{eq:swparam} is based on simulations in which $v_{\rm w}$ is not too close to $1$. However, in the model that we consider, $\gamma_{\rm w}$ is typically at least $\mathcal{O}\left(1\right)$ in thermal transitions since the bubble walls only reach a terminal velocity due to the loop suppressed $\gamma_{\rm w}$ dependent friction. Meanwhile, in tunnelling transitions the terminal value of $\gamma_{\rm w}$ can be huge and the sound wave contribution to the gravitational wave spectrum is only significant if this is reached. This difference in dynamical regimes introduces some unavoidable uncertainty into our analysis. Directly studying systems with extremely large $\gamma_{\rm w}$ appears impossible in simulations, however further theoretical developments might be possible and such progress could potentially be combined with extrapolations of results from simulations.\footnote{We also note that the amplitude of the fit that we use differs from that in \cite{Caprini:2015zlo}, which was based on earlier simulations.}

The parameter $\kappa_{\rm sw}$ in Eq.~\eqref{eq:swparam} is the proportion of the vacuum energy transferred to kinetic energy in the plasma. If the bubble walls reach a constant velocity, $\kappa_{\rm sw}$ can be determined from a hydrodynamical analysis of the wall and plasma system. Since this depends only on the properties of the hidden sector, we can adapt results calculated for visible sector phase transitions \cite{Espinosa:2010hh}, and if $v_{\rm w} \simeq 1$
\beq
\kappa_{\rm sw} \simeq \frac{\alpha_{\rm h}}{0.73 + 0.083 \sqrt{\alpha_{\rm h}}+ \alpha_{\rm h}} ~.
\eeq
In the model we consider $\alpha_h$ is typically $\sim 1$ in thermal transitions, which corresponds to efficient conversion to kinetic energy in the plasma and little energy going into directly heating it. In tunnelling transitions, $\alpha_h$ is $\gg 1$, so if the bubble walls reach a terminal velocity long before colliding then $\kappa_{\rm sw}$ is basically 1. As argued, if the bubble walls are still accelerating when they collide, the energy transfer to the plasma is negligible and we can simply set $\kappa_{\rm sw} =0$. In the intermediate regime for which the bubble walls reach a terminal velocity not long before colliding, we can estimate 
\beq \label{eq:kswdef}
\kappa_{\rm sw} = 1 - \gamma_*/\gamma_0~,
\eeq
which also takes the correct values in the other regimes.

\subsection{Gravitational waves signals} \label{GW-spectra}

We are now ready to study the gravitational wave signal produced by a particular phase transition, and analyse its detectability in future experiments \cite{TheLIGOScientific:2016dpb,TheLIGOScientific:2014jea,Caprini:2015zlo,Sathyaprakash:2012jk,Carilli:2004nx, 2010CQGra..27h4013H}. To do this we use the standard sensitivity curves corresponding to the noise power spectral density (see \cite{Thrane:2013oya}) which are widely used in the phase transition literature.

First we note that the amplitude of a gravitational wave signal is strongly suppressed if $\alpha \ll 1$, regardless of which source dominates, and only models with relatively large $\alpha$ have a chance of being observed. In Section~\ref{sec:phaset} we saw that $\beta/H_* \sim 10$ in tunnelling transitions and $\beta/H_* \sim 10$ -- $100$ in thermal transitions, regardless of whether the hidden sector is cold or at the same temperature as the visible sector. Combined with Eqs.~\eqref{eq:peakfreqenv} and \eqref{eq:peaksound}, this means that a signal's peak frequency is always parameterically set by the Hubble parameter at the time of the transition. Tunnelling transitions usually have a slightly smaller peak frequency than thermal transitions for a given value of the Hubble parameter, due to having smaller $\beta/H_*$. However, there are likely to exist models with thermal transitions for which $\beta/H_* \simeq 10$ as well, for example due to a nucleation rate that is only weakly temperature dependent, so this is not a sharp prediction.

In thermal transitions $\alpha \simeq \epsilon^3 \alpha_{\rm h}$, where $0.1 \gtrsim \alpha_{\rm h} \gtrsim 10$ in typical models (c.f. Eq.~\eqref{eq:alphaepi}). As discussed in Section~\ref{subsec:friction}, the bubble wall velocity is always finite in thermal transitions in the model that we consider, so the vast majority of the energy is transferred to the plasma. The gravitational wave spectrum is therefore dominantly produced by sound waves, and has a high frequncy fall off $\propto f^{-3}$. The amplitude of the gravitational wave signal produced by bubble collisions is suppressed by 
\beq
\kappa_{\rm c} \simeq  \gamma_{\rm w} \sigma / (R_* \rho_{\rm vac})~,
\eeq
and is negligible as expected.

\begin{figure}[t]\begin{center}
\includegraphics[height=8cm]{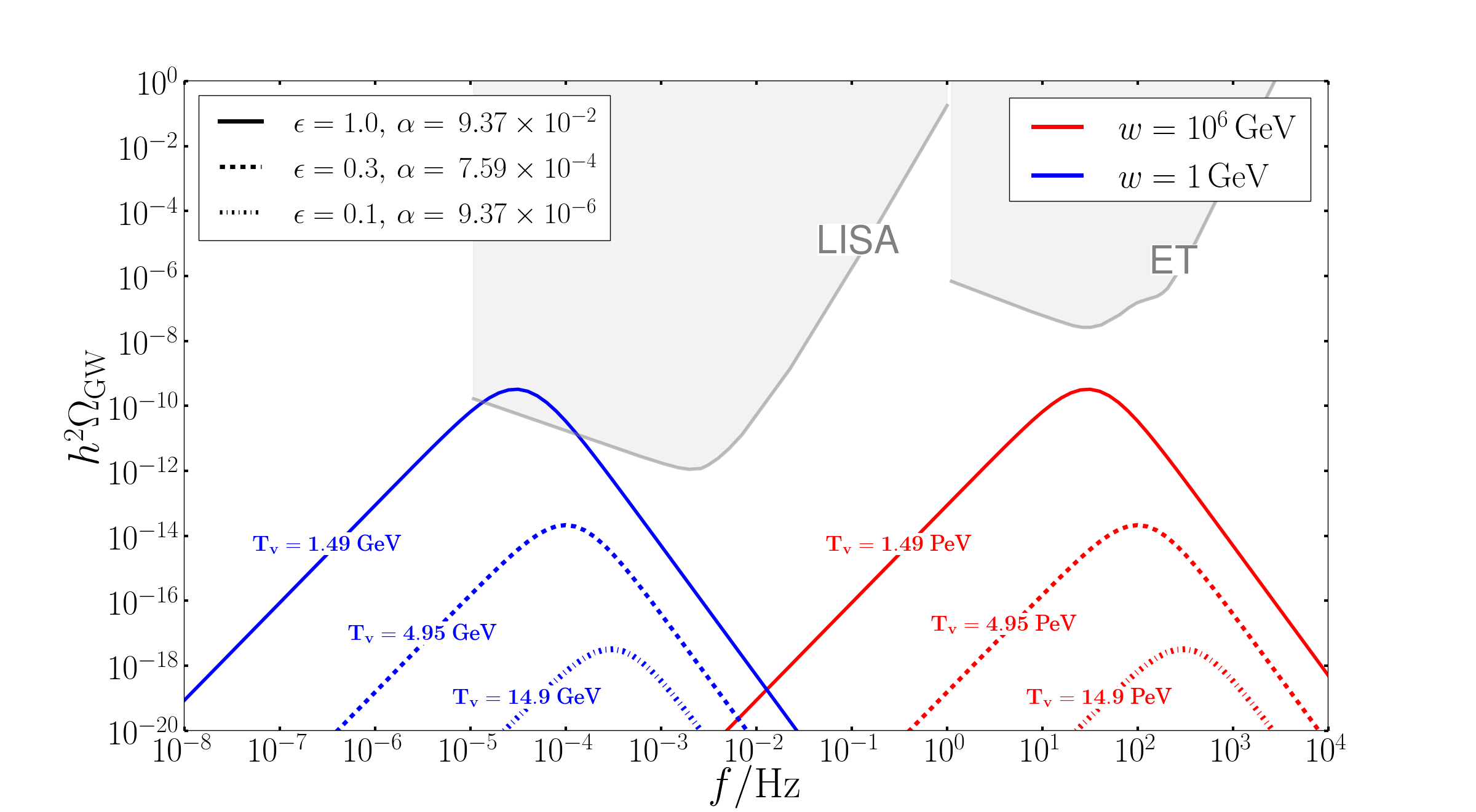}
\end{center}
\caption{Example gravitational wave spectra from thermal transitions in the model described in Section~\ref{sec:model}. Results are shown for different hidden and visible sector temperature ratios $\epsilon$, and scales $w$ which determine the hidden sector temperature that the transition occurs at. We fix $g=2$, and adjust $\tilde{m}^2$ such that $\alpha_h=1$ remains fixed. As $\epsilon$ decreases, the thermal energy in the visible sector is increasingly greater than the energy released by the phase transition. Thus with decreasing $\epsilon$ the amplitude of the gravitational wave spectrum becomes heavily supressed and rapidly moves out of sensitivity of any upcoming searches. Somewhat larger values of $\alpha_{\rm h} \lesssim 50$ are possible in some parts of parameter space, which would increase the detection possibilities slightly.}
\label{fig:thermalTspec}
\end{figure}

In Figure~\ref{fig:thermalTspec} we plot example spectra from thermal transitions with different values of $\epsilon$, and fixed $\alpha_{\rm h}= 1$ (so that $\alpha$ varies). We see that, in this frequency range, observable signals are only possible in such models for $\epsilon \simeq 1$, corresponding to a hidden sector at almost the same temperature as the visible sector prior to the transition. The closeness of the hidden and visible sector temperatures required for a detectable signal is slightly relaxed for larger $\alpha_{\rm h}$, and values $\alpha_{\rm h} \simeq 50$ are possible in parts of the hidden sector parameter space that we consider, corresponding to significant super cooling. However, the minimum $\epsilon$ that leads to an observable signal only scales approximately as $\propto \alpha_{\rm h}^{1/3}$, so sectors with parametrically small $\epsilon$ remain unobservable even in this case.

Unlike in thermal transitions, the gravitational wave signal from a tunnelling transition can be dominated either by emission from sound waves or from bubbles collisions, depending on whether the bubble walls reach a terminal velocity, or not, respectively. The boundary between the two regimes is determined by Eq.~\eqref{eq:gammara}, via the factors $\kappa_{\phi}$ and $\kappa_{\rm sw}$ in Eqs.~\eqref{eq:kphidef} and \eqref{eq:kswdef} (we study the change in spectral shape moving between these two regimes shortly). Regardless of which dominates, the amplitude of the signal is again strongly suppressed if $\alpha \ll 1$. 

\begin{figure}[t]\begin{center}
\includegraphics[height=8cm]{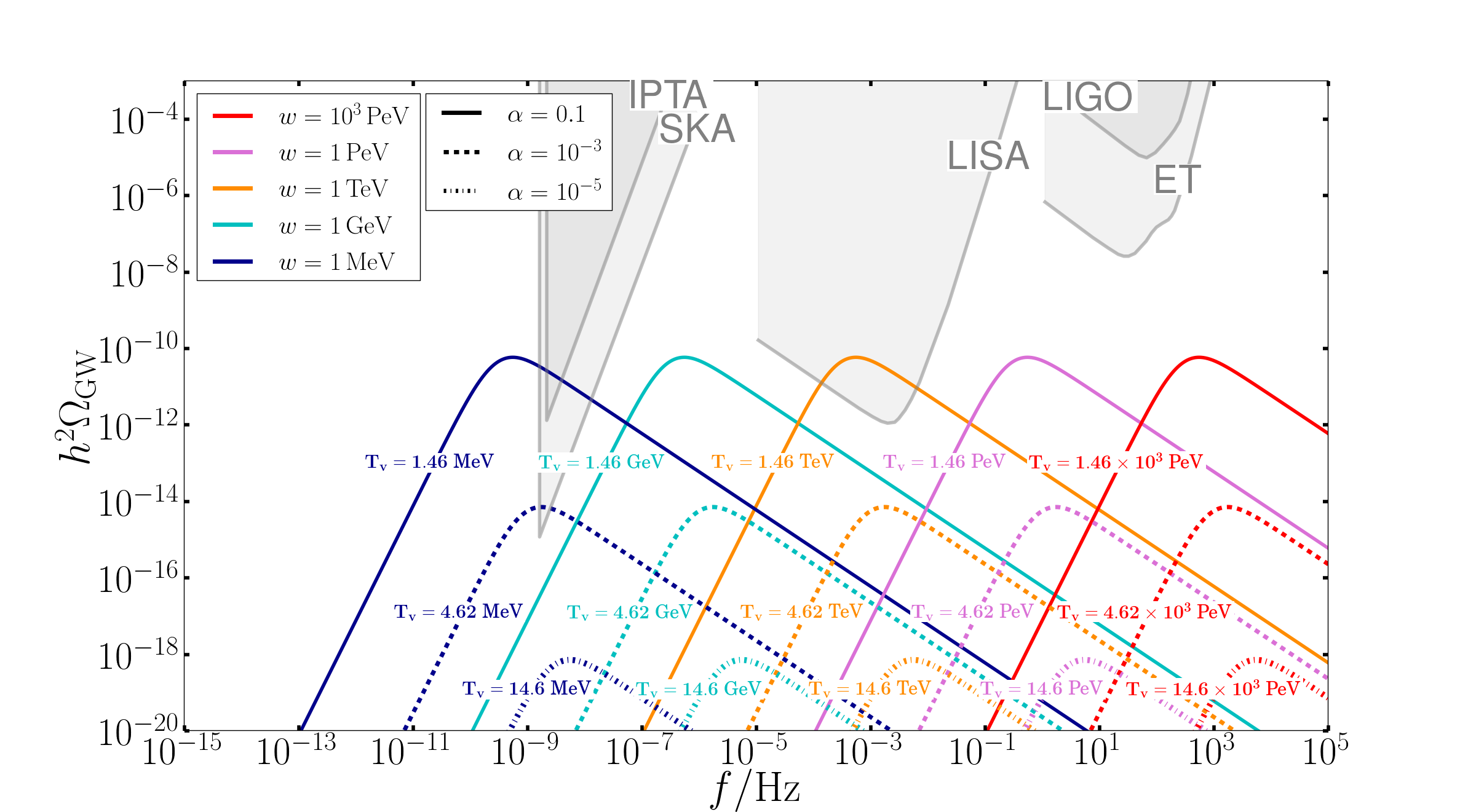} \end{center}
\caption{Examples of the gravitational wave spectra emitted by tunnelling transitions in the absence of friction (due to the hidden sector being extremely cold, with $\epsilon \approx 0$). We show results for different values of the relative energy released by the transition compared to that in the visible sector: $\alpha$, as well as for different hidden sector scales $w$, and we fix $g=2$. For a fixed $\alpha$ the scale $w$ determines the visible sector temperature at the time of the transition. In such models the bubbles walls runaway and gravitational waves are dominantly produced by the collision of bubbles. A signal is only observable for models in which $\alpha$ is relatively large.}
\label{fig:tunnelspec}
\end{figure}

The spectrum from a tunnelling transition in the absence of any hidden sector thermal bath, i.e. if $\epsilon = 0$, is shown for different values of $\alpha$ in Figure~\ref{fig:tunnelspec}. This corresponds to all the released energy going into the bubble walls, and the signal from a model with any $\epsilon$ that is significantly less than the boundary value between effective runaway and terminal wall velocities will also look almost identical. It can be seen that only transitions for which $\alpha \gtrsim 10^{-4}$ have a chance of being detected in currently proposed experiments, and this is also the case for tunnelling transitions with larger $\epsilon$ for which the signal comes dominantly from sound waves. In tunnelling transitions $\alpha$ varies from $\sim \epsilon^3$ to $\sim 1$, with a roughly logarithmic distribution of models in this range, so detectable signals only occur in relatively small parts of parameter space.

The amplitude of a gravitational wave signal from a phase transition in a sector at the same temperature as the visible sector can easily vary by orders of magnitude across different models. For example, a transition with relatively small bubble walls speeds, significantly below the speed of light, will produce a strongly suppressed signal. Hence, observation of a signal with a small amplitude would not be sufficient to claim discovery of a sector that is not thermalised with the visible sector. Similarly the peak frequency of the gravitational wave signal from a sector at the same temperature as the visible sector can be adjusted by altering the scale of the hidden sector. This flexibility in the signals from thermalised sectors seems unpromising for discriminating between thermalised and non-thermalised hidden sectors. However, we now describe two possibilities that might provide an insight into the source of a gravitational wave signal, were one to be discovered.

\subsubsection*{Runaway vs non-runaway transitions}

The high frequency fall off of a measured gravitational wave spectrum could give a clear indication that it arose from a transition in which the bubble walls runaway with negligible energy transfer to the plasma, rather than a transition in which a significant proportion of the released energy is transferred to the plasma. The spectrum emitted by the former will have a fall off $\propto f^{-1}$ -- $f^{-1.5} $ since it is dominantly produced by bubble collisions  (allowing for the uncertainty between simulations and theoretical predictions), which is significantly different to the $\propto f^{-3}$ fall off in transitions with finite bubble wall speeds.\footnote{As mentioned previously, we assume that the contribution from turbulence is sub-leading.} 

\begin{figure}[t]\begin{center}
\includegraphics[height=8cm]{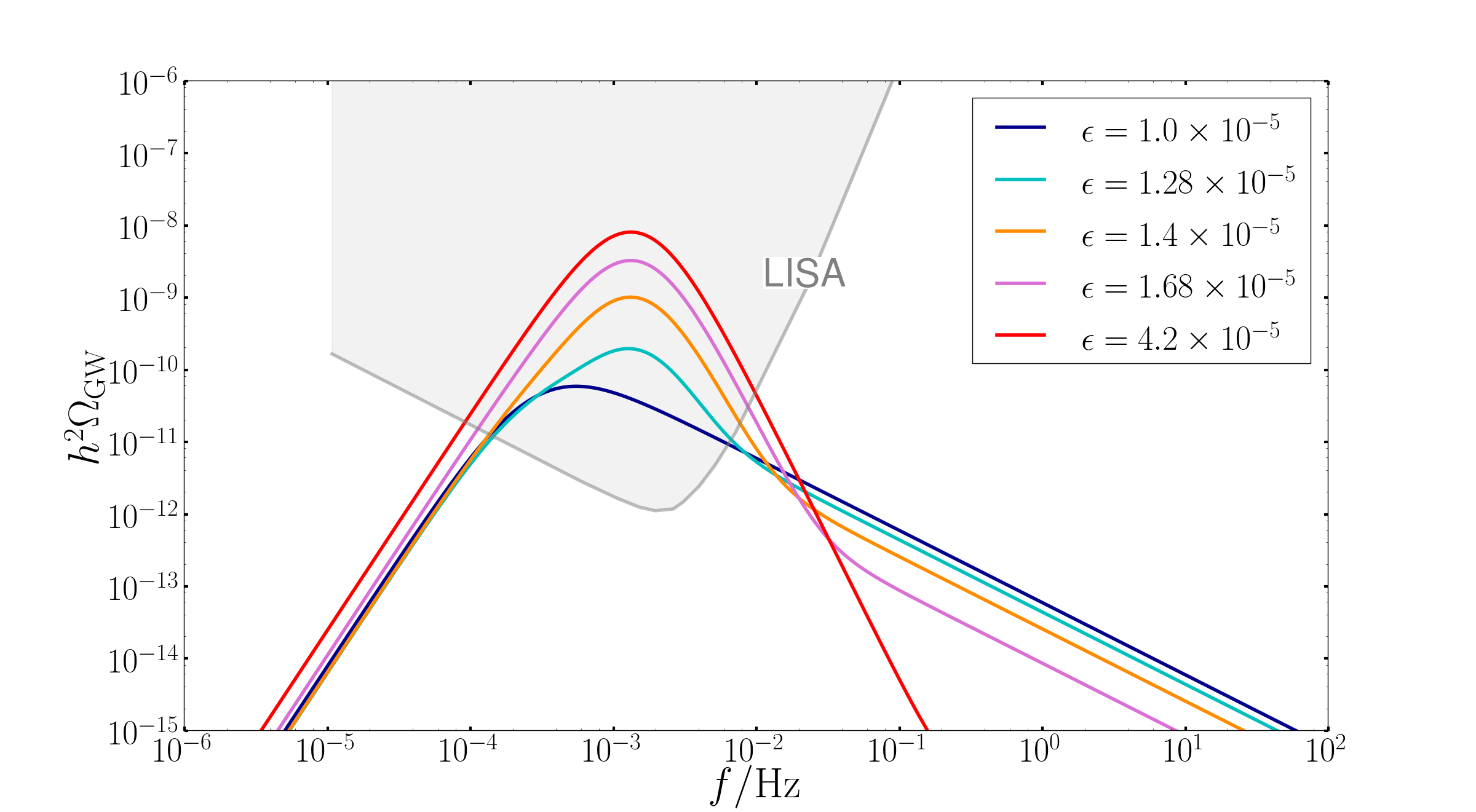}\end{center}
\caption{The gravitational wave spectrum emitted by tunnelling transitions for different values of the hidden sector temperature (via different $\epsilon$). We fix the visible sector temperature at the time of transition $T_{\rm v *} = 1461 ~ \GeV$, $w=10^3 ~\GeV$,  $\alpha=0.1$, and $g=2$ across the different spectra. As $\epsilon$ is increased from $1 \times 10^{-5}$ to $4.2 \times 10^{-5}$, the hidden sector temperature at the transition increases, so the friction on the bubble walls increases. For $\epsilon \gtrsim  1.2\times 10^{-5}$, the bubble walls reach a terminal velocity before they collide, and the gravitational wave spectrum goes from being dominated by bubble collisions (with a $\propto f^{-1}$ high frequency fall off) to being dominated by sound waves (with a $\propto f^{-3}$ fall off).}
\label{fig:epsitun}
\end{figure}

A transition with negligible energy transfer to the plasma can arise in a cold hidden sector that goes through a tunnelling transition at a time when the number density of the hidden sector thermal bath is sufficiently small. In Figure~\ref{fig:epsitun} we show the change in the gravitation wave spectrum emitted by a tunnelling transition in the hidden sector that we consider, for different values of $\epsilon$. As $\epsilon$ increases, the friction grows and the bubble walls reach their terminal velocity prior to collisions. The fast drop in $\kappa_{\rm c}$ after $\epsilon$ passes the boundary between regimes, Eq.~\eqref{eq:epraw}, results in a sharp transition in the signal's shape.\footnote{These results are calculated using fits from simulations with much smaller values of $\gamma_{\rm w}$ than will occur in the regime plotted, which introduces some uncertainty. However, based on theoretical models, we still expect a significantly less steep fall off of the spectrum in a bubble dominated collision.}

As discussed, it is unclear if runaway bubbles are possible in thermal transitions in models without gauge bosons. Even if this can occur, at least $\simeq 10\%$ of the released energy is transferred to the plasma through the $\gamma_{\rm w}$ independent friction in typical models. The amplitude of the gravitational wave signal from sound waves Eq.~\eqref{eq:swparam} appears to be larger than that from bubble collisions, fit by Eq.~\eqref{eq:scalaransatz}, in the limit $v_{\rm w} \simeq 1$, and the peak frequencies of the signals from sound waves and bubble collisions differ by a factor $\simeq 2$. Therefore, it seems plausible that a sound wave contribution could usually be distinguished in most such cases. It would be interesting to study the shape of the spectra produced in this case in more detail, especially if further results from large scale numerical simulations become available.\footnote{Additionally, an intermediate $f^{-1}$ dependence could arise in transitions with $v_{\rm w}$ close to the speed of sound \cite{Hindmarsh:2017gnf}, confirming the need for careful analysis of the spectral shape after initial discovery for models to be discriminated.}

This difference in gravitational wave spectral shape between a runaway tunnelling, and a non-runaway tunnelling or non-runaway thermal transition could be experimentally distinguished, provided that the signal is detected over a reasonably wide frequency range. Although it may not be possible to exclude a runaway thermal transition that is completely dominated by bubble walls, the unusual model building requirements for this to arise mean that detection of such a spectral shape would still be extremely interesting. From Figure~\ref{fig:epsitun} we see that a $f^{-1}$ fall off, which would be enough to show that emission from bubble walls dominates, could be observed at e.g. LISA if $\alpha$ is close to $\mathcal{O}\left(1\right)$. Discovery of a signal would also prompt further targeted experimental investigation with increased sensitivity, which would allow for more detailed analysis of the frequency dependence of the spectrum.

\subsubsection*{Phase transitions around BBN}

Another potentially observable possibility is that a thermal transition in a hidden sector that is slightly colder than the visible sector, or a tunnelling transition, could lead to a detectable gravitational wave signal in a frequency range that is not possible from a hidden sector at the same temperature as the visible sector, due to cosmological constraints.\footnote{This has also been carefully considered in \cite{Breitbach:2018ddu}, which was posted on the arXiv as we were preparing our manuscript. Our results are compatible with theirs, and the reader is refereed there for an alternative very nice discussion of this possibility.}

The constraint on the effective number of relativistic degrees of freedom in the hidden sector at the time of BBN, Eq.~\eqref{bbn-bound-1}, means that a phase transition at a temperature $T_{{\rm v}*} \lesssim 10~\MeV$ is not possible in a hidden sector at the same temperature as the visible sector. The corresponding peak frequencies of gravitational wave spectra emitted at around such temperatures are (assuming sound wave domination, and $v_{\rm w} \simeq 1$)
\beq \label{eq:peakbbn}
  f_\mathrm{sw} = 6.1 \times 10^{-8} ~ \mathrm{Hz} \, 
   \left( \frac{\beta/H_*}{100} \right) \left( \frac{T_{v*}}{10 ~\MeV}\right)~,
\eeq
where $\beta/H_*$ is typically $50$ -- $100$, but might plausibly be as small as $10$ in some models. As discussed at the end of Section~\ref{sec:cosmo} transitions $T_{{\rm v}*} \gtrsim 10~\MeV$ are possible in hidden sectors at the same temperature as the visible sector, although the parameter space is quite strongly constrained unless the transition is at a much higher temperature  $T_{{\rm v}*} \gg 10~\MeV$.

In contrast, a cold hidden sector can have a transition at any visible sector temperature, and therefore emit gravitational waves with any peak frequency. It needs only satisfy the constraints, studied in Section~\ref{sec:cosmo}, on its energy density ($\epsilon< 0.2$ and $\alpha < 0.015$) and contain light hidden sector states that avoid the universe being over closed, none of which restrict the allowed time of a transition.\footnote{We impose the slightly strong constraints on $\epsilon$ and $\alpha$ from the CMB rather than BBN so that the light hidden sector states can definitely evade the relic abundance bound, however the difference is only small.} 

\begin{figure}[t]\begin{center}
\includegraphics[height=5.34cm]{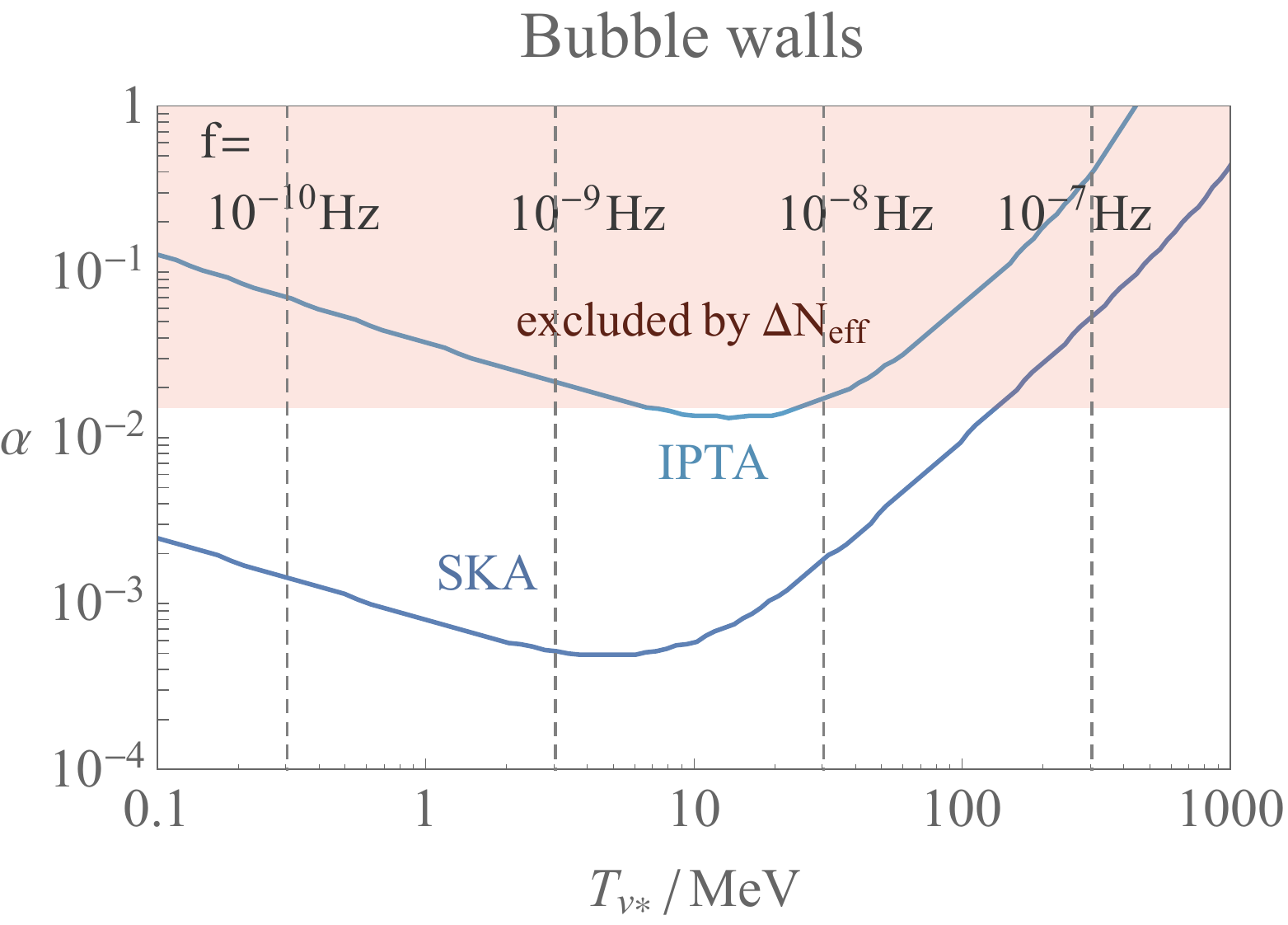}\hspace{0.05cm}
\includegraphics[height=5.34cm]{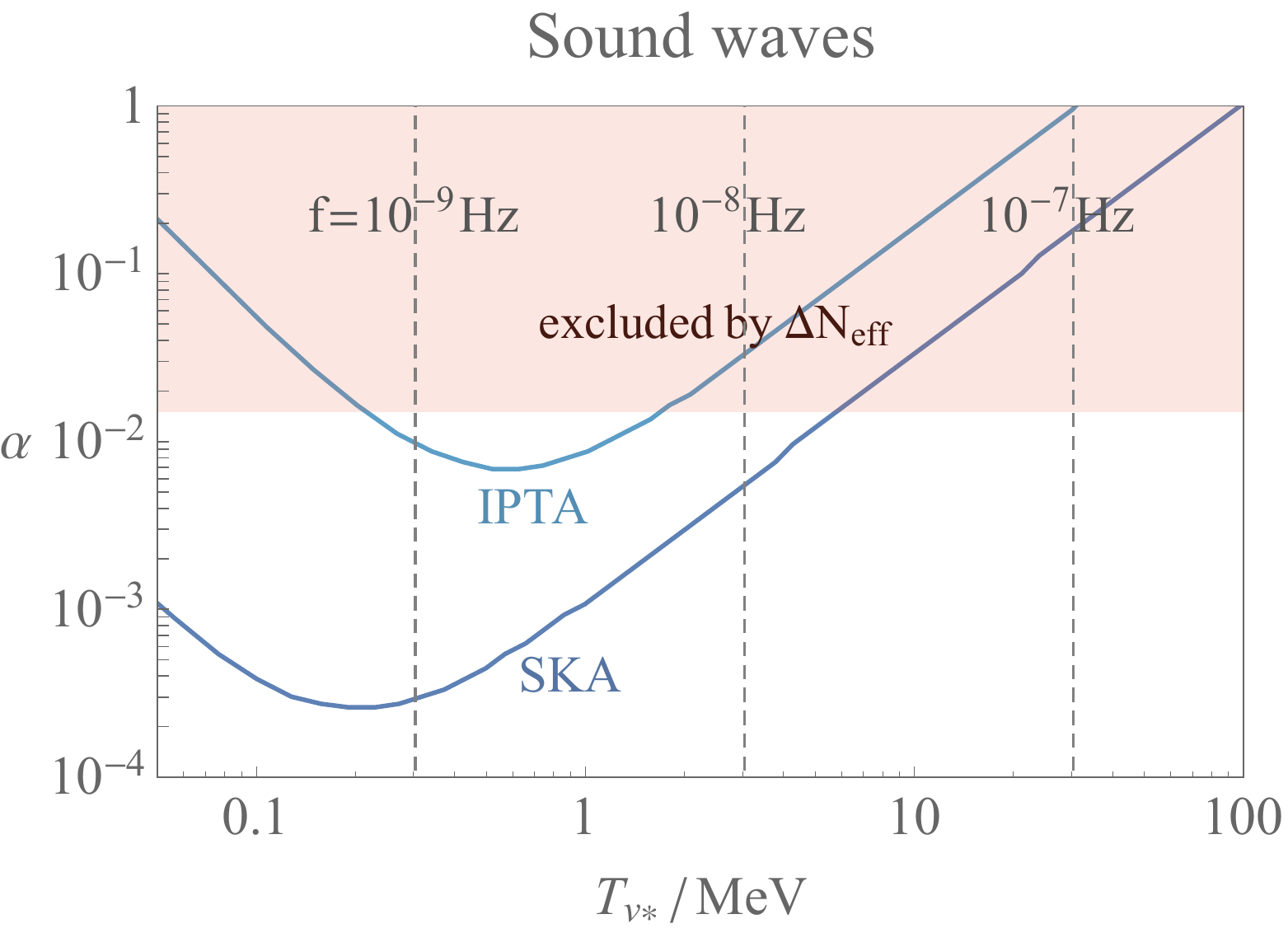} \end{center}
\caption{The minimum value of $\alpha$ that leads to a gravitational wave signals that can be detected at SKA and IPTA as a function of the visible sector temperature at the time of the transition $T_{\rm v *}$. Results are shown for runaway bubble walls in an effective vacuum with $\beta/H_*=10$ (left) and sound waves from a phase transition that is strong (but without runaway bubble walls) with $\beta/H_*=100$ (right). The corresponding peak frequencies of the emitted gravitational wave signal are also indicated. We assume the hidden sector states decay to light hidden sector states, and the effects of these on the number of relativistic degrees of freedom constrains $\alpha < 0.015$. Once parameterised in terms of $\beta/H_*$, $\alpha$, and $T_{{\rm v}*}$ the results are independent of the particular hidden sector considered. Constraints from BBN mean we do not expect to see a signal from a hidden sector at the same temperature as the visible sector for $T_{v*}\lesssim 10$ MeV, which means that a signal with a peak frequencies $\lesssim 10^{-9}~{\rm Hz}$ would be a strong indication of a signal originating in a cold hidden sector (even allowing for freedom in $\beta/H_*$). }
\label{fig:bbntimes}
\end{figure}

In Figure~\ref{fig:bbntimes} we plot the minimum values of $\alpha$ that could be detected in the upcoming experiments IPTA and SKA, for runaway tunnelling transitions and non-runaway thermal transitions with $v_w \simeq 1$. The results are shown as a function of the visible sector temperature at the time of the transition, and the corresponding peak frequencies of the signals are also given. For our purposes a model is detectible if any part of its gravitational wave spectrum crosses the sensitivity curve of an experiment.

 In these plots we assume $\beta/H_* = 10$ and $\beta/H_* = 100$ for the tunnelling and thermal cases respectively, although the results are not too sensitive to these choices.
 We also show the constraint on $\alpha$ from CMB observations, assuming that all of the energy released by the phase transition ends up in light hidden sector states (the previously mentioned model dependent opening up of the allowed values of $\alpha$ at $T_{v*} \gtrsim 10~\MeV$ is not shown). Although the hidden sector that we have studied provides an example of a model that can lead to the signals we consider here, the results in Figure~\ref{fig:bbntimes} are general to any cold sector, once its phase transition is parameterised in terms of $\alpha$ and $\beta/H_*$ (and the dominant source of gravitational waves is determined).

We see that there are significant regions of parameter space in which a cold hidden sector has a phase transition at a visible sector temperature $T_{{\rm v}*} \lesssim 10~\MeV$ that leads to a gravitational wave signal that is detectable at SKA, without being excluded by cosmological constraints. There is also a small region of parameter space in which the same is true of IPTA. 
In these parts of parameter space the peak frequencies of the gravitational wave signals are in the range $\lesssim 10^{-9}~{\rm Hz}$ and $\lesssim 10^{-8}~{\rm Hz}$ for runaway bubbles and sound waves respectively. 

There is no sharp lower bound on the peak frequency of a gravitational wave signal from a hidden sector at the same temperature as the visible sector, since there is some freedom in the value of $\beta/H_*$, and the fit Eq.~\eqref{eq:Ssw} may not be precisely correct for all transitions. However, it seems extremely unlikely that such a sector would produce a signal with a peak frequency $\lesssim 10^{-9}~{\rm Hz}$. 

As a result, the discovery of a signal with a sufficiently small peak frequency would be a strong sign of the existence of a cold hidden sector that was out of thermal equilibrium with the visible sector. The first observation of a signal is likely to be through integration over frequencies, which might not be sufficient to exclude the possibility that the observed signal is the low frequency tail of a spectrum peaked at higher frequencies. However, with further observation, combined with much more careful analysis of experimental uncertainties than we attempt, the location of the spectrum's peak could be reliably determined.

\section{Conclusions}\label{sec:conc}

We have studied phase transitions in hidden sectors that are cold relative to the visible sector, and the possible resulting gravitational wave signals. Using a simple example model we have shown that phase transitions in extremely cold hidden sectors can proceed through nucleation of bubbles by tunnelling, as opposed to via thermal fluctuations as occurs if a hidden sector is at roughly the same temperature as the visible sector. This is the case both for high and low scale reheating.

The gravitational wave signals of a tunnelling transition in a cold hidden sectors can be observably different to those produced by a typical thermal transition in a warm hidden sector, owing to the differing dynamics of the bubble walls in the two cases. In particular, in a sufficiently cold hidden sector there is no significant bath of hidden sector states, which if present transfer a proportion of the energy released by the phase transition to bulk motions of the hidden sector plasma.\footnote{At present, it is unknown if thermal transitions in unusual models could also lead to overwhelmingly bubble collision sourced spectra, and this is an important area for future work. However, even if such thermal transitions turn out to be possible, the theories that give rise to them will have to have unusual properties, so observation of a bubble collision dominated signal will remain exciting, even if it does not conclusively point to a cold hidden sector.} In this regard, a key difference between our results and earlier work (e.g. \cite{Caprini:2015zlo}) that considered the possibility of gravitational wave signals dominated by bubble collisions, is that  we include the recently discovered $\gamma_{\rm w}$ dependent friction  \cite{Bodeker:2017cim}. This prevents runaway walls in typical thermal transitions, even in those with significant supercooling (which is sufficient to render the $\gamma_{\rm w}$ independent friction negligible). 

We have also seen that, due to BBN constraints, cold hidden sectors can lead to gravitational wave signals in a frequency range that would otherwise not be possible. These two effects mean that if a gravitational wave spectrum were to be discovered, it might be possible to show that it comes from a sector that is cold relative to the visible sector. Given that such a sector would be far out of reach of direct exploration by experiments, this would be an extremely interesting discovery.

More pessimistically, gravitational wave signals from cold hidden sectors are strongly suppressed in thermally nucleated transitions if the hidden sector is extremely cold, and likewise in tunnelling transitions if the energy released is much smaller than the energy in the visible sector. Observable signals therefore only arise in relatively small regions of hidden sector parameter space.

Although the literature contains extensive careful work studying the dynamics of phase transitions and the gravitational waves that they emit, there are a number of key theoretical questions that are still open, in addition to those already discussed. In particular, there is significant uncertainty on the friction experienced by bubble walls. It is known that thermal transitions in which gauge bosons change mass cannot lead to runaway bubble walls, however precise expressions for the friction are unknown. The extrapolation of the gravitational wave spectrum determined in simulations to the physically relevant scale separation is also challenging. Finally, no reliable numerical determination of the contribution of turbulence to the gravitational wave spectrum has been carried out. Further developments in all of these directions would be extremely useful for obtaining sharper predictions of the gravitational wave signals expected from different types of hidden sectors.

For our present work we have used a simple perturbative hidden sector model to demonstrate the features of interest. Although we believe that this captures many of the physical possibilities, it would certainly be interesting to also consider other forms of hidden sector in detail. From a UV perspective it is perhaps most reasonable to expect that a hidden sector phase transition will happen due to a gauge theory running into strong coupling, since hidden sectors containing non-Abelian gauge groups appear to be ubiquitous in many classes of string compactifications. It would therefore be worthwhile to attempt to study the phase transitions, and resulting gravitational wave signals, from such hidden sectors in detail. This is likely to be challenging due to the uncertainties involved in finite temperature lattice field theory, although progress may be possible.\footnote{For some strongly coupled hidden sectors theoretical arguments, e.g. epsilon expansions or considerations of the central symmetries, can sometimes provide additional insight. However, it seems unlikely that these could provide all the information about the dynamics of a phase transition.} 

It has also been proposed that phase transitions could happen in the early universe through vacuum decay in string theory constructions \cite{GarciaGarcia:2016xgv}. While we have focused on field theory models of the hidden sectors, we expect that the gravitational wave signals in such string theory scenarios will most closely resemble those from tunnelling transitions in an extremely cold hidden sector, which as we have seen, are dominated by emission from bubble collisions. Similarly to tunnelling transitions, the high frequency fall off of such a spectrum would be a strong indication that the signal did not simply come from a typical thermally nucleated transition in a field theory sector.

Although we have seen that observable signals can occur, our results perhaps make the possibility of detecting a gravitational wave signal from one of a multitude of very cold hidden sectors appear less likely. Tunnelling transitions can lead to observable signals even in extremely cold hidden sectors, however this requires a coincidence between the bubble nucleation rate, which varies exponentially with the model's parameters, and the energy density of the false vacuum, in order that $\alpha$ is fairly close to $\mathcal{O}\left(1\right)$. Furthermore, we have shown that in some parts of parameter space a hidden sector can make a viable cosmological history impossible even if it is extremely cold and has no impact on BBN or CMB observations, regardless of how the cosmological constant is tuned.

To conclude, gravitational wave signals remain a compelling possibility for discovering otherwise extremely hard to detect hidden sectors. Reliably determining the map between the properties of a hidden sector and the gravitational wave spectrum that it emits is therefore an important task, especially in light of the likely future improvements in experimental sensitivities. In our present work we have made a step towards this goal, however much remains to be done.

\section*{Acknowledgements}
MF  and  AW  are  funded  by  the  European  Research  Council  under  the  European Union’s Horizon 2020 programme (ERC Grant Agreement no.648680 DARKHORIZONS).  In addition, the work of MF was supported partly by the STFC Grant ST/P000258/1.

\bibliographystyle{JHEP}
\bibliography{reference}

\appendix

\section{Maintaining a temperature hierarchy}
\label{app:maintaing-temp}

A necessary condition to maintain a temperature difference between the hidden and visible sector is that they stay out of thermal equilibrium with each other. This is the case if the (temperature dependent) interaction rate between them, $\Gamma_I$, satisfies
\beq \label{eq:GammaI}
\Gamma_I < H ~,
\eeq
at all times between reheating and the hidden sector phase transition. 
The form of $\Gamma_I$ depends on which of the hidden and visible sectors is hotter. We assume that the visible sector is hotter, in which case Eq.~\eqref{eq:GammaI} becomes
\beq \label{eq:nsv}
\left< \sigma v \right> n_{\rm v} < H(T_{\rm v}) ~,
\eeq
where $\left< \sigma v \right>$ is the thermally averaged cross section between the visible and hidden sectors and $n_{\rm v}$ is the number density of the relevant states in the visible sector.\footnote{A hidden sector that is hotter than the visible sector is less interesting. The universe must be dominated by the visible sector before BBN, so in this case the energy density in the hidden sector must be transferred to the visible sector before this time. There are regions of parameter space in which this is possible. However the gravitational wave signal that would arise is similar in shape and amplitude to if the two sectors were in thermal equilibrium throughout.}

For relatively small temperature differences the condition in Eq.~\eqref{eq:GammaI} is approximately sufficient to maintain a hierarchy. However, for large temperature ratios a stronger condition is required. The hidden sector temperature must not be increased by energy transfer from the visible sector, even if this  energy is not enough for the two sectors to reach the same temperature. The resulting constraint can be estimated by demanding that the energy density transferred per Hubble time to the hidden sector is smaller than that already present in the hidden sector, all times prior to the phase transition. This corresponds to
\beq \label{eq:cond}
\frac{n_{\rm v}^2 \left<\sigma v\right> T_{\rm v}}{H^4} \lesssim  \frac{T_{\rm h}^4}{H^3} ~,
\eeq
where we have dropped factors of order 1. If a hidden sector is initially colder than this, but the portal coupling is not large enough for full thermalisation, it would be partially heated up until Eq.~\eqref{eq:cond} is satisfied.

The hidden sector that we consider can couple to the visible sector in multiple different ways, and we focus on a simple Higgs portal operator as an example. This interaction takes the form
\beq \label{eq:hpop2}
\mathcal{L} \supset -\frac{1}{2}\lambda_p \left|\Phi\right|^2 \left|H \right|^2 ~,
\eeq
where $H$ is the SM Higgs doublet. Although $\lambda_p =0$ is technically natural, it is not unreasonable to suppose a non-zero value might be present, for example due to heavy states that couple to both sectors. 

If the hidden sector phase transition happens when the visible sector temperature is above the scale of the electroweak (EW) phase transition, the cross section between the two sectors at the relevant times is
\beq
\left<\sigma v\right> \sim \frac{\lambda_p^2}{32\pi T_{\rm v}^2} ~,
\eeq
and Eq.~\eqref{eq:cond} becomes
\beq \label{eq:cond2}
\lambda_p \lesssim 10^{-8} \epsilon^{3/2} \left( \frac{w}{\GeV} \right)^{1/2}~.
\eeq
When $\epsilon=1$ this matches the previously known condition for the two sectors to not thermalise at temperatures above the EW scale $\lambda\lesssim 10^{-7}$ \cite{Bento:2001yk}. As expected, 
smaller values of the portal couplings are required to maintain a large temperature hierarchy between the two sectors, and tiny portal couplings are needed if $\epsilon \ll 1$. 

The condition on $\lambda_p$ is different if the hidden sector phase transition happens when the visible sector temperature is below the EW scale. In this case the dominant energy transfer from the visible sector happens immediately after the EW phase transition, since at this point the decay channel $h \rightarrow \phi\phi$ is open and there is still a thermal population of the SM Higgs (which will later be exponentially suppressed).\footnote{Since we assume that the hidden sector is colder than the visible sector this is automatically kinematically allowed.} The resulting energy transfer can be approximated from the Higgs branching fraction to the hidden sector \cite{Bento:2001yk}, which leads to the constraint to maintain a temperature hierarchy
\beq \label{eq:cond1a}
\lambda_p \lesssim 10^{-10} \epsilon^2 ~.
\eeq
This again matches the condition for thermalisation when $\epsilon =1$, and it is much stronger for small $\epsilon$. Such tiny portal couplings are far beyond the reach of any direct experimental searches.\footnote{A temperature difference could be maintained with larger values of the portal coupling 
if the reheating temperature of the universe was below the EW scale. In this case energy is only transferred to the hidden sector through scattering of light SM fermions via an off-shell Higgs, and the rate that this occurs at is strongly suppressed. Portal couplings that are large enough to have observational consequences might be possible in this case, although we do not investigate it further.}

\section{Transitions after low scale reheating in the hidden sector}
\label{app:lowreheat}

In this Appendix we analyse the impact of relaxing our assumption in Section~\ref{sec:phaset} that the hidden sector is reheated above the temperature at which its high temperature phase is favoured. In doing so we show that a tunelling transition still only occurs if the hidden sector is cold relative to the visible sector.

The hidden sector might be reheated to a temperature that is below the scale at which its symmetry is restored, or it may not be reheated at all. If the hidden sector is in the symmetry preserving phase at the end of inflation, this can result in a phase transition happening through tunnelling rather than thermal fluctuations. Such initial conditions arise most naturally if the Hubble parameter during inflation is above the symmetry restoration scale. We will see that this scenario is easily possible in models with a significant temperature difference between the hidden and visible sectors, but it cannot occur for hidden sectors at the same temperature as the visible sector if the scale of the phase transition is $\lesssim \TeV$.\footnote{It may be possible to arrange for these initial conditions despite a lower Hubble scale during inflation, evading our argument, by introducing a more complex prior cosmological history. However this simply postpones the question of the initial conditions to earlier times.}

We suppose that the Hubble scale during inflation $H_I > w$, and for the moment we also assume that the hidden and visible sectors are at the same temperature. The visible sector reheat temperature must be $\gtrsim 5~\MeV$ to preserve the successful predictions of BBN. This constrains the inflaton decay rate $\Gamma_{\rm inf}$ to
\beq
\MeV \lesssim \left( \Gamma_{\rm inf} M_{\rm Pl} \right)^{1/2} ~,
\eeq
which means that the maximum temperature after inflation $T_{\rm max}$ (which is larger than the reheating temperature) will be \cite{Giudice:2000ex}
\beq
\begin{aligned}
T_{\rm max} &\simeq \left(H_I M_{\rm Pl}^2 \Gamma_{\rm inf} \right)^{1/4} \\  
&\gtrsim 3000 \GeV \left(\frac{H_I}{100~\GeV} \right)^{1/4} ~.
\end{aligned}
\eeq
Therefore for hidden sectors at a scale $w\lesssim \TeV$ the universe automatically reaches a temperature at which the hidden sector symmetry is restored if $H_I \gtrsim w$. Subsequently, as the temperature drops a thermal transition will happen in preference to a tunnelling transition, as in the previously considered case.\footnote{A minor caveat to this argument is that the relation between the Hubble parameter and temperature is altered during reheating, because the universe is matter dominated. This makes it slightly less likely that a transition completes through thermal fluctuations. In practice the difference between the two actions, e.g. from Eqs.~\eqref{eq:s3scal} and \eqref{eq:s4scal}, is sufficiently large that this does not lead to a tunnelling transition in any of the parameter space of our model.}

For hidden sectors at higher scales it is possible that its symmetry is restored during inflation and the sector subsequently undergoes a tunnelling transition, despite being at the same temperature as the visible sector. This requires $T_{\rm max} < w$, and that the tunnelling transition happens 
before $H$ drops below $\sim w^2/M_{\rm Pl}$, when the universe would reenter an inflationary phase. The conflicting pressures of reheating above the scale of BBN and having $T_{\rm max} < w$  mean that tunnelling transitions only happen in a small region of parameter space. Additionally, the entropy injection by the inflaton decays after the phase transition dilute the present day gravitational wave signal in this case.

These issues are avoided if the hidden sector is cold relative to the visible sector. The Hubble scale during inflation can be sufficiently high that the hidden sector symmetry is restored, $H >w$. However, if the inflaton decays dominantly to the visible sector the hidden sector temperature might never get close to $w$, despite the visible sector being reheated above the scale of BBN. In this case a thermal transition will not take place, but provided $S_4$ is sufficiently large a tunnelling transition can happen before the hidden sector vacuum energy density dominates the universe. 

The condition that a tunnelling transition completes is $\Gamma_4 \gtrsim \sqrt{\rho_{\rm vac}}/\left(3 M_{\rm Pl} \right)$, as before. The parts of parameter space that satisfy this condition are those in Figure~\ref{fig:typeofT} right that undergo a transition (including parts shown as going through a thermal transition in the previous context of that plot). The visible sector temperature at the time of the transition is again given by Eq.~\eqref{eq:TvisTun} (assuming, for simplicity, that the transition happens after reheating completes i.e. when the Hubble parameter is $\lesssim H\left(T_{\rm RH}\right)$).

\section{Gravitational waves signals}\label{app:GW}

Here we summarise results from the literature that support the parameterised fits of the gravitational waves signals that we use in Section~\ref{sec:gvsignals}, and our neglect of emission from turbulence and field oscillations. In particular, we argue that despite uncertainty on the details of the signals, the prediction of a significantly steeper fall off in the spectrum from sound waves compared to that from bubble collisions is robust.

Theoretical uncertainty arises because results from numerical simulations require extrapolation to the large scale separations in the physical phase transitions. In particular, there is a large difference between the scale of the microscopic physics fixing the critical bubble radius and bubble wall thickness, and that of the Hubble distance setting the typical bubble radius at collision. Meanwhile theoretical predictions require simplifying approximations. Consequently, a combination of these approaches is required to maximise the reliability of predictions.

\subsection{Bubble collisions}

The envelope approximation has been used to make theoretical predictions of frequency dependence of the gravitational wave spectrum emitted by bubble collisions. The low frequency part of the gravitational wave spectrum is fixed by causality to have a power law $h^2 \Omega_{\rm GW}\left(f\right) \propto f^{3}$ (where $f$ is the frequency), since its source is uncorrelated on time scales bigger than $1/\beta$. As early as 1992, Kosowsky et al. proposed that the high frequency tail follows a power law $f^{-1.5}$. This work has since been complemented with an independent analytical estimate in \cite{Jinno:2016vai} that predicts a $f^{-1}$ dependence. The subsequent inclusion of a slight time dependence of the bubble nucleation rate, as might be the case in a thermal transition, only altered this prediction by $\mathcal{O}$(10\%) and seems to leave the peak magnitude of the signal unchanged \cite{Jinno:2017ixd}.

Numerical simulations using the envelope approximation were performed in \cite{Huber:2008hg} and support the idea of a gradual high frequency fall-off, finding a result that is closer to $f^{-1}$. Subsequent work in \cite{Weir:2016tov}, also using the envelope approximation, obtained compatible results, and argues that this is independent of the bubble nucleation rate. 

More recently, facilitated by advances in computational power and techniques, \cite{Cutting:2018tjt} carried out large simulations of the gravitational wave emission in vacuum phase transition, by directly evolving a scalar field with a suitable potential. This allowed the validity of the envelope approximation itself to be tested precisely. They found a signal with an $f^{-1.5}$ high frequency decay and an overall magnitude that is the same as is obtained using the envelope approximation to within a factor of a few.\footnote{The earlier direct simulation of a scalar field \cite{2012JCAP...10..001C} finds a gravitational wave signal with an amplitude dramatically smaller than the theoretical prediction, however this was not confirmed by \cite{Cutting:2018tjt}, and the origin of this result remains unclear.}  We will for the moment put this potentially important result to one side, and work with the established results from the envelope approximation, in the knowledge that the precise magnitude and spectrum of our expected signal might change slightly.

\subsection{Sound waves}

The gravitational wave power spectrum from acoustic waves has been studied in increasingly large simulations, in a range of hydrodynamic regimes and with different bubble wall velocities, in \cite{Hindmarsh:2013xza,Hindmarsh:2015qta,Hindmarsh:2017gnf}. Even with the limited scale separations accessible in simulations, the sound waves are found to be an important source of gravitational waves if a significant fluid component exists, and that emission continues for at least a Hubble time after the bubbles collide. Following this numerical progress, \cite{Hindmarsh:2016lnk} and \cite{Jinno:2017fby} proposed analytical models for the emission of gravitational waves from sound waves.

\cite{Hindmarsh:2015qta} also discusses the extrapolation to the physical regime in the scenario that the bubble walls reach a terminal velocity and emission by sound waves overwhelmingly dominates that from bubble collisions. The model in \cite{Hindmarsh:2016lnk} was also tested on the lattice in \cite{Hindmarsh:2017gnf} and the two descriptions were found to be in broad agreement.

\subsection{Turbulence in the plasma}

The collision of acoustic sound shells stirs the plasma, and potentially produces turbulent flows. These start at the scale of the average bubble radius $R_*$, and cascade to smaller scales, until damped by viscosity \cite{Kosowsky:2001xp,Kahniashvili:2008pf,Kahniashvili:2008pe,Kalaydzhyan:2014wca}. Gravitational waves are emitted during this process, potentially over a period of several Hubble times. If the hidden sector plasma is ionised with respect to a light gauge field, the evolution of the system is more complex. In this case hidden sector magnetic fields, sourced by the phase transition, are amplified dramatically and feed in to the dynamics, leading to a magneto-hydrodynamical (MHD) system \cite{Caprini:2006jb,Caprini:2009yp,Kahniashvili:2009mf}.

Studying turbulence with numerical simulations is challenging, owing to the characteristic wide range of scales in the problem. The largest available simulations of gravitational wave production from phase transition, \cite{Hindmarsh:2015qta,Hindmarsh:2017gnf}, find no evidence for a significant turbulent contribution to the gravitational wave spectrum in a thermal transition, in the case that MHD effects are not included. However, these simulations are of relatively weak transitions. Numerical results analysing systems with larger fluid velocities, for which turbulence is expected to be more important, would be very welcome in determining the impact of this contribution (although existing simulations are already a numerical tour de force, and involve substantial computational resources).

When considering phase transitions in the visible sector, for which the plasma is ionised and MHD is the correct description of the system, the spectrum of gravitational waves emitted by turbulence is often assumed to take the standard Kolmogorov form, with an analytically calculated amplitude \cite{Caprini:2009yp}. This can be parameterised as \cite{Caprini:2015zlo}
\beq \label{eq:mhdparam}
h^2 \Omega_\text{turb}(f) =
3.35 \times 10^{-4} \left(\frac{H_*}{\beta}\right) \left(\frac{\kappa_\text{turb} \alpha}{1 +
    \alpha} \right)^2
    \left(\frac{100}{g_*} \right)^{\frac{1}{3}}
v_\mathrm{w}  \, S_\text{sw}(f)
\eeq
where $\kappa_\text{turb}$ is the fraction of latent heat released in the phase transition that is converted into turbulent flows. The spectral shape is
\begin{equation}
   S_\text{turb}(f) = \left(\frac{f}{f_\text{turb}}\right)^3  \left(
   \frac{1}{[1 + (f/f_\text{turb})]^{11/3}(1 + 8\pi f/h_*) } \right)^{7/2} ~,
\end{equation}
where 
\beq
h_* = 1.65 \times 10^{-3}~\mu\mathrm{Hz} \left(\frac{T_{{\rm v}*}}{100~\GeV}\right) \left(\frac{g_{{\rm v}*}}{100}\right)^{1/6} ~,
\eeq
and the approximate peak frequency is
\begin{equation} \label{eq:peakmhd}
  f_\mathrm{turb} = 2.7\times 10^{-2} \, \mu\mathrm{Hz} \, 
  \frac{1}{v_\mathrm{w}} \left( \frac{\beta}{H_*} \right) \left( \frac{T_{v*}}{100 \,
    \mathrm{GeV}}\right) \left( \frac{g_{{\rm v}*}}{100} \right)^\frac{1}{6}
\end{equation}
The fraction of the total energy density available for turbulent flows is parameterised as $\kappa_{\text{turb}}= \epsilon_{\text{turb}} \kappa_{\rm sw} $, where $\epsilon_{\text{turb}}$ accounts for the efficiency with which bulk motion is converted to vorticity. It is generally thought that $\epsilon_{\text{turb}} \simeq 0.05$ -- $0.1$, although this is not conclusive.

For our present purposes, we assume that the contribution to the gravitational wave spectrum from turbulence is subleading to that from sound waves, which is accurate if the fits in Eq.~\eqref{eq:swparam} and \eqref{eq:mhdparam} are accurate. If this is the case, the gravitational wave emission by turbulence will not affect any of the observational features that we subsequently identify, and can be safely neglected. Such an assumption is however not fully justified for the model that we consider. As discussed, emission from the plasma will only be significant if the bubble walls reach a terminal speed, but the value of $\gamma_{\rm w}$ in such cases is still typically large. Despite this, we use the existing fits as a reasonable first approximation, which could be improved in future work.

\subsection{Long lived oscillations after collisions}

When bubbles collide they produce large amplitude non-linear oscillations in the scalar field, which act as a relatively long lasting source of gravitational waves. These were observed in \cite{2012JCAP...10..001C}, and studied carefully in \cite{Cutting:2018tjt} using large numerical simulations. This component of the spectrum  is clearly identifiable in simulations of vacuum transitions, however it appears to be subsumed in (or simply subdominant to) the sound wave emission in simulations of thermal transitions.

The gravitational wave spectrum produced in long lived oscillations (in simulations of vacuum transitions) has a peak frequency that appears to be $\sim 1/l_0$, where $l_0$ is the thickness of the bubble walls, which has the parametric form \cite{Cutting:2018tjt}
\beq
l_0 \simeq \frac{1}{\sqrt{d^2V/d\phi^2}} ~.
\eeq
The limited scale separations achievable in simulations means that the low frequency tail contribution from this source is mixed into the spectrum of gravitational waves from bubble collisions. At the physically relevant scale separation between $H^{-1}_*$ and $l_0$, this emission would presumably lead to an extremely high frequency peak in the gravitational wave spectrum, well separated from the signal from bubble collisions. 

In the largest simulations carried out so far, the gravitational wave spectrum from this source is found to still be growing when the simulations are ended \cite{Cutting:2018tjt}, and at this time their contribution to the energy in gravitational waves is starting to exceed that from bubble collisions. However, as argued in \cite{Cutting:2018tjt}, this appears to be an artifact of the limited scale separations in simulations. By analysing the rate at which the spectrum grows in simulations, and using the physical expectation that oscillations only persist for a time $H_*^{-1}$, they predict that the relative energy in the gravtiational waves produced by such oscillations $\Omega_{\rm osc}$ compared to that from bubble collisions scales as
\beq \label{eq:oscrel}
\frac{\Omega_{\rm osc}}{\Omega_{\rm coll}} \simeq 10^{2} \frac{l_0^2}{H_*^2}~.
\eeq

In the physically relevant regime, Eq.~\eqref{eq:oscrel} predicts, very approximately, $\Omega_{\rm osc}/\Omega_{\rm coll} \lesssim w^2/M_{\rm Pl}^2$ (dropping numerical factors, gauge couplings, and factors of $\beta/H_*$). Therefore, in transitions happening in an effective vacuum, Eq.~\eqref{eq:oscrel} means that a completely negligible fraction of the total energy density in gravitational waves comes from long lived field oscillations sourced this way\footnote{If further investigation reveals that the amplitude of the signal from this source is larger than predicted by Eq.~\eqref{eq:oscrel}, it would be interesting to investigate if it could lead to observable high frequency signals in some models.} The suppression in transitions with finite bubble walls speeds will be even stronger, since the vast majority of the energy released by the transition goes directly into the plasma. As a result, we do not need to consider this source when we subsequently study the gravitational wave spectra from hidden sectors.

\end{document}